 \documentclass[preprint2,longabstract]{aastex}

\def\heao1{{\it HEAO-1\/}}

\def\iras{{\it IRAS\/}}
\def\spitzer{{\it Spitzer\/}}

\def\herschel{{\it Herschel\/}}

\def\iso{{\it ISO\/}}
\def\herschel{{\it Herschel\/}}
\def\iras{{\it IRAS\/}}

\newcommand{\ltsima}{$\; \buildrel < \over \sim \;$}
\newcommand{\simlt}{\lower.5ex\hbox{\ltsima}}
\newcommand{\gtsima}{$\; \buildrel > \over \sim \;$}
\newcommand{\simgt}{\lower.5ex\hbox{\gtsima}}

\def\gsimeq{{_>\atop^{\sim}}}
\def\lsimeq{{_<\atop^{\sim}}}
\def\lesssim{\mathrel{\hbox{\rlap{\hbox{\lower4pt\hbox{$\sim$}}}\hbox{$<$}}}}
\def\gtrsim{\mathrel{\hbox{\rlap{\hbox{\lower4pt\hbox{$\sim$}}}\hbox{$>$}}}}

\shorttitle{Are $z>2$ Herschel galaxies proto-spheroids?}
\shortauthors{Pozzi et al.}

\begin{document}



\title{Are the bulk of $z>2$ Herschel galaxies proto-spheroids?}


\author{F.Pozzi\altaffilmark{1}}
\affil{Dipartimento di Fisica e Astronomia, Universit\`a degli Studi di Bologna,
Viale Berti Pichat 6/2, I--40127 Bologna, Italy}
\email{f.pozzi@unibo.it}

\author{F. Calura and C. Gruppioni}
\affil{INAF --- Osservatorio Astronomico di Bologna, Via Ranzani 1, I--40127
Bologna, Italy}

\author{G. L. Granato}
\affil{INAF --- Osservatorio Astronomico di Trieste, Via G. B. Tiepolo
11, I--34131, Trieste, Italy}

\author{G. Cresci}
\affil{INAF --- Osservatorio Astronomico di Arcetri, Via Largo Enrico
Fermi 5, I--50125 Firenze, Italy}

\author{L. Silva}
\affil{INAF --- Osservatorio Astronomico di Trieste, Via G. B. Tiepolo
11, I--34131, Trieste, Italy}

\author{L. Pozzetti}
\affil{INAF --- Osservatorio Astronomico di Bologna, Via Ranzani 1, I--40127
Bologna, Italy}

\author{F. Matteucci\altaffilmark{2}}
\affil{Dipartimento di Fisica, Universit\`a degli Studi di Trieste,
Via Valerio 2, I--34127 Bologna, Italy}

\and

\author{G. Zamorani}
\affil{INAF --- Osservatorio Astronomico di Bologna, Via Ranzani 1, I--40127
Bologna, Italy}


\altaffiltext{1}{INAF --- Osservatorio Astronomico di Bologna, Via Ranzani 1, I--40127
Bologna, Italy.}

\altaffiltext{2}{INAF --- Osservatorio Astronomico di Trieste, Via G. B. Tiepolo
11, I--34131, Trieste, Italy}


\begin{abstract}

We present a backward approach for the interpretation of the evolution of the
near-infrared and the far-infrared luminosity functions
across the redshift range $0<z<3$. In our method, late-type galaxies are treated by means of a parametric phenomenological method
based on PEP/HerMES data up to $z{\sim}$4, whereas spheroids are described by 
means of a physically motivated backward model. 
The spectral evolution of spheroids is modelled by means of a single-mass model, associated to 
a present-day elliptical with $K-$band luminosity comparable to the break of the local early-type 
luminosity function. The formation of proto-spheroids is assumed to occurr across the redshift range $1\le z \le 5$.
The key parameter is represented by the redshift $z_{0.5}$ at which half proto-spheroids are already formed. A statistical study indicates 
for this parameter values between $z_{0.5}=1.5$ and $z_{0.5}=3$. We assume as fiducial value $z_{0.5} \sim 2$, 
and show that this assumption allows us to describe accourately the redshift distributions and the source counts. By assuming $z_{0.5} \sim 2$ at the
far-IR flux limit of the PEP-COSMOS survey, the PEP-selected sources observed 
at $z>2$ can be explained as progenitors of local spheroids caught during their formation.
We also test the effects of mass downsizing by dividing the spheroids into three populations of different present-day stellar masses. 
The results obtained in this case confirm the validity of our approach, i.e. that the bulk of proto-spheroids can be modelled by means of a single model which describes the evolution of galaxies at the break of the present-day early type $K$-band LF.

\end{abstract}




\section{Introduction}
\label{intro_sec}

Achieving a complete understanding of the origin of the local
dichotomy of spheroids and disc galaxies has been one of
the main objectives in astrophysics during the past years,
as well as obtaining an accurate measure of the star formation history over
cosmic time.

From the pioneering \iras\ satellite results in the local Universe (e.g. \citealt{1987ARA&A..25..187S}) and the detection of a cosmic
infrared background as energetic as the optical/near-IR background
(e.g. \citealt{1996A&A...308L...5P}), it is now well established that most
energy radiated by newly formed stars is heavily absorbed by dust and re-emitted in the infrared (IR)
band. In the last decade, the \iso\ and the \spitzer\ satellite individually detected
IR sources up to $z{\sim}$1 (e.g. \citealt{1999A&A...351L..37E}, \citealt{2002MNRAS.335..831G}) and
$z{\sim}2$ (e.g. \citealt{2004ApJS..154...70P}, \citealt{2008AJ....135.1050S}) in
the mid-IR band, but their capabilities at far-IR wavelengths (i.e. where the dust
reprocessed emission peaks) were still strongly limited by source
confusion.

Nowadays, the {\it Herschel Space Observatory} (\citealt{2010A&A...518L...1P}) has allowed to properly
measure the IR luminosity function of galaxies up to $z{\sim}$4 (\citealt{2013MNRAS.432...23G}[hereafter GPR13], see also
\citealt{2013A&A...553A.132M}), thanks to its mirror of
3.5-m and an observing spectral range between 60 and 670~$\mu$m.  The derived IR luminosity density
(${\rho}_{IR}$) confirms the  \spitzer\  24-$\mu$m based results (e.g. \citealt{2007ApJ...660...97C},
\citealt{2009A&A...496...57M}, \citealt{2010A&A...515A...8R}) up to
$z{\sim}2$, revealing that the IR luminosity density increases steeply from $z=0$ up to
$z{\sim}$1, then flattens between $z{\sim}$1 and $z{\sim}$3, to
decrease at $z>$3.

In the recent paper by \cite{2013A&A...554A..70B}, the IR
{\it Herschel} luminosity function derived by GPR13 has been
combined with the luminosity function in the far-UV from
\cite{2012A&A...539A..31C}, in order to achieve an estimate of the redshift evolution
of the total (far-UV + IR) star formation rate density (SFRD). The SFRD
is always dominated by the IR emission, whereas
the UV contribution increases steeply from $z=0$ up to $z{\sim}2.5$, where it flattens
and settles on a plateau up to the highest redshift sampled by the survey ($z\sim 3.6$).
This suggests that the peak of the dust
attenuation, occurring at $z{\sim}1$, is delayed with respect to the
SFRD plateau ($z\sim {2-3}$), derived from the far-UV.

The accurate determination of the star-formation history up to $z{\sim}$4
has rendered particularly urgent the issue of explaining theoretically this behaviour,
as well as understanding how the observed SFRD evolution is linked to the galaxy formation process.

Within the current `concordance' cosmological paradigm, which  sees
a $\Lambda$-Cold dark Matter ($\Lambda$CDM)-dominated universe,
the formation of structures is hierarchical, since small
dark matter (DM) halos are the first to collapse, then interact and merge
to assemble into larger halos
(e.g. \citealt{2008MNRAS.385.1155L}, \citealt{2009MNRAS.392..553F}).

 In cosmological $\Lambda$CDM-based semi-analytical models (SAMs) for galaxy formation,
the most uncertain assumptions concern the behaviour of the baryonic matter.
In the first classical SAMs implementations, baryonic matter was assumed to follow the DM in all the interaction and
merging processes and spheroid galaxies were formed from several merging episodes
of smaller sub-units, with the most massive galaxies as the last
systems to assemble. 
Indeed, more recent results suggest that spheroids may form over a
wide redshift range by means of both galaxy mergers and disk
instabilities (e.g. \citealt{2011MNRAS.414.1439D}; \citealt{2014arXiv1407.0594P}).  
In particular, the latter process should be of fundamental importance
if the vast majority of $z{\approx}2$ quasi-stellar
objects is associated with unperturbed
systems, as indicated by recent results (e.g. \citealt{2012MNRAS.425L..61S}). 

  Although accounting for the recent updatings, SAM galaxy
 formation models still encounter difficulties to reconcile properly with the well-established
anti-hierachical evolution of spheroids, a feature which is
known as ``downsizing'' and which is supported by a
large amount of observational data (\citealt{1994A&A...288...57M};
\citealt{1996AJ....112..839C}, \citealt{2004MNRAS.350.1005K}, but see also \citealt{2009ApJ...701.1765M}).

Moreover, $\Lambda$CDM models highlight difficulties for reproducing 
 the evolution of IR galaxies (in particular the bright-end IR
 luminosity function at high-$z$, see e.g. \citealt{2011MNRAS.416.2962F},
\citealt{2012MNRAS.421.1539N}, Gruppioni, Calura, Pozzi et al. in prep.).
The basic problem of the
dominance of a merger-driven evolution is the duration of the
star-formation in spheroids,
which extends over times much longer than those indicated by the basic stellar
population diagnostics of local ellipticals (\citealt{1994A&A...288...57M}; \citealt{2006ARA&A..44..141R}).

Doubts on the dominant role of mergers in driving the star formation histories of galaxies
is also supported by the observation of systems with regular kinematics and showing little
signs of major mergers (\citealt{2009ApJ...697..115C}; \citealt{2009ApJ...706.1364F}; \citealt{2011A&A...528A..88G}), as well as the tight scaling
relations
of star-forming galaxies, such as the mass-SFR plane (\citealt{2011ApJ...739L..40R}) and the
fundamental metallicity relation (\citealt{2010MNRAS.408.2115M}; \citealt{2010A&A...521L..53L}), apparently incompatible with the
stochastic nature of galaxy mergers (see e.g. \citealt{2008ApJ...687...59G}).

In \cite{2009Natur.457..451D}, a new
scenario is proposed where galaxy formation is not mainly merger-driven,
but occurs instead via a `cold stream accretion' process,
which represents the dominant mode for galaxy formation.
Such a scenario is able to account for several
galaxies properties at high-$z$ (i.e. gas-richness and extended rotationally
supported disks). In the proposed model, the streams responsible for
the creation of rotationally-supported discs could also supply the
turbulence necessary for the formation of spheroids.

\citet{2009MNRAS.397..534C}, building upon the model by
\citet{2004ApJ...600..580G}, proposed an alternative scenario for the origin of the
local dichotomy between spheroids and discs in galaxies, in the context of the
hierarchical build up of dark matter (DM) haloes. Noting that
the growth of DM haloes occurs in two phases
(e.g. \citealt{2003MNRAS.339...12Z}), namely a fast collapse one,
featuring a few major merger events,
followed by a late, quiescent accretion onto the halo outskirts,
they proposed that the first and second phases
of halo growth drive two distinct modes for the evolution of baryonic
matter, favoring the development of the spheroidal and disc components of galaxies,
respectively. As a result, in that semi-analytic model, the spheroidal
component develops rapidly and at early times ($z \gtrsim 2$), mimicking the
classic `monolithic model' (e.g. \citealt{1976MNRAS.176...31L};
\citealt{1994A&A...288...57M}.)

Although a similar picture can
account for the duality between spheroids and discs,
obtaining a full, theoretical understanding of the basic
properties of galaxies as traced by current multi-wavelength surveys
over cosmic time, is still a challenging task.

At present, the lack of a complete theoretical picture for galaxy formation encouraged
several attempts to reproduce the basic observational
properties of galaxies over cosmic time via `hybrid' models,
which generally match simple physical models to parametric
approaches. A few attempts (i.e. \citealt{2005MNRAS.357.1295S};
\citealt{2011ApJ...742...24L}; \citealt{2013ApJ...768...21C}, \citealt{2014MNRAS.438.2547B})
have been able to provide a reasonable description
of the multi-wavelength properties of galaxies, from the near-IR up to sub-mm wavelengths.
However, the treatment of the chemical evolution of dust grains in such systems has often been simplistic,
i.e. assuming a constant carbon-to-silicate ratio or dust-to-gas ratios
simply scaling with metallicity.

In the work of \cite{2009MNRAS.394.2001S} detailed predictions on the evolution of the dust chemical composition in
galaxies of different morphological types have been matched to the spectro-photometric code
GRASIL (\citealt{1998ApJ...509..103S}), which takes into account dust reprocessing,
to analyse the evolution of the resulting spectral energy
distributions (SEDs). \cite{2009MNRAS.394.2001S} showed that a
detailed treatment of the dust chemical composition is particularly important when the proto-spheroids SEDs are taken into
account, since dust assumptions tuned to match the dust depletion pattern of the Milky Way
turned out as particularly inadequate to describe their observables.

In this paper, we present a new backward phenomenological approach where the spectro-photometric evolution of dust in proto-spheroids is
calculated self-consistently, i.e. from idealised star formation histories able to account for a large set
of observational data at various redshifts, including their dust
content and local abundances ratios (\citealt{2008A&A...479..669C} [hereafter CPM08]; \citealt{2011A&A...525A..61P}) and the evolution of their SEDs (\citealt{2009MNRAS.394.2001S}).
Beside our physically motivated `backwards' model for the evolution of proto-spheroids,
the evolution of late-types galaxies is also taken into account
through a parametric, phenomenological approach.

Our aim is to reproduce a variety of
multi-wavelength observables, including source counts, redshift distributions and luminosity functions
at various redshifts and from the near-IR to the far-IR.

The basic questions we aim to address
are to which extent the star-forming galaxies
detected by \herschel\ at high-$z$, could be considered as the progenitors of local
early galaxies (i.e. proto-spheroids), and which quantitative constraints can be achieved from
current observations regarding the dominant epoch for spheroid formation.

The paper is structured as follows: in Section 2 we describe how the
model deals with the late type and proto-spheroids populations, in
Section 3 we compare the far-IR and $K$-band observables
with the model
predictions, and in Section 4 we discuss the formation redshift and
mass buildup of the proto-spheroids population.

All magnitudes reported in this paper are in the AB system (\citealt{1983ApJ...266..713O}).
Throughout this paper, we adopt a flat ${\Lambda}CDM$ cosmology with
$H_{0}=71$ kms$^{-1}$, $\Omega_{m}=0.27$ e $\Omega_{\Lambda}=0.73$.
 
\section{Model description}
\label{model_sec}

\begin{figure}
\centering
\includegraphics[width=80mm,height=75mm]{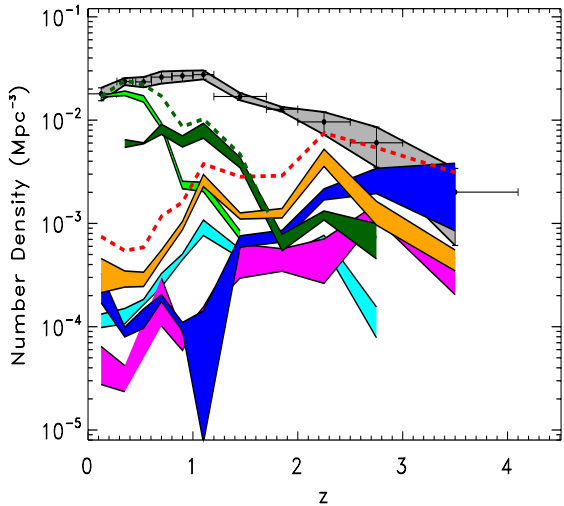}
\caption{Evolution of the comoving number density of the PEP sources
  up to $z{\sim}4$ (taken from Fig. 18 of GPR13). Black filled
  dots with error-bars within the ${\pm}1{\sigma}$ uncertainty
  region (grey filled area) represent the total sample.
  The coloured shaded areas represent the contribution from
  different populations, highlighted as follows:
  green:
  spiral; dark-green: SF-AGN(Spiral); cyan: starburst; orange:
  SF-AGN(SB); magenta: AGN2; blue: AGN1.
  The dark-green dashed line corresponds to
  the sum of the spiral+SF-AGN(Spiral) populations; the red
  dashed line corresponds to the sum of the
  starburst+SF-AGN(SB)+AGN2+AGN1 populations. }
\label{number_density_fig}
\end{figure}


The basic idea behind our approach is similar to the one of previous
`hybrid' models  used to interpret far-IR and optical data
(see \citealt{2005MNRAS.357.1295S}; \citealt{2011ApJ...742...24L}; \citealt{2013ApJ...768...21C}), where
galaxies are divided into two main classes, i.e. early-type and
late-type galaxies, and each class treated in a different way.

The evolution of late-type
galaxies is followed by means of a parametric, classical `backward' approach, which consists in
evolving a local luminosity function (LLF) back with redshift (\citealt{2011MNRAS.416...70G}).
The observables are reproduced by evolving the LLF
of different populations,
each characterized by a set of Spectral
Energy Distributions (SEDs), in density and luminosity.
This approach has been traditionally used to
interpret far-IR source counts (e.g. \citealt{2004ApJ...609..122P},
\citealt{2009ApJ...701.1814V}, \citealt{2010A&A...517A..74F},
\citealt{2011MNRAS.416...70G}, \citealt{2011A&A...529A...4B}).

The evolution of proto-spheroids is followed by means of a
model in which the chemical composition of dust is calculated in detail
and self-consistently on the basis of its star formation history.
In this model, the evolution of the SED of the proto-spheroid is
followed as a function of time.

The main reason for developing this approach derives from the
need of finding a model able to account for
data in various bands.
Up to now, infrared backward
evolutionary models, although accurately accounting for
the dust far-IR emission (linked to the ongoing
star-formation, hence to young stellar populations), have neglected the need to account for other observables
related to evolved stellar populations, such as the $K$-band
data, which trace the oldest stellar populations.
Data from $K$-band up to far-IR
 wavelengths can be simultaneously reproduced only if a self-consistent
model that accounts for the time-dependent galaxy SED is considered.

The description of our method can be subdivided into three steps. First, we divide far-IR sources into late-type
galaxies and `potential' proto-spheroids on the basis of their SED and, more
importantly, on the basis of the evolution (positive or negative) of their comoving number density as
a function of redshift $z$ (see
Sect. \ref{populations_sec}). This approach is different from that used in other works, such as \cite{2013ApJ...768...21C}, where IR galaxies are
classified according to their
redshift, with high-$z$ ($z$$>$1.5) star-forming sources generally interpreted
as proto-spheroids, and sources at lower-$z$ ($z$$<$1.5)
 as late-type galaxies.
Secondly, we normalize the proto-spheroid
galaxy number density according to the local $K$-band LF. Finally, we use our physical, chemo-spectrophotometric
SED evolution (\citealt{2009MNRAS.394.2001S}) to calculate the source counts and redshift distributions for
the proto-spheroid population.
The proto-spheroids formation
 redshift is left as the only free parameter;
for this quantity, we explore an appropriate range
in order to simultaneously reproduce the $K$-band and far-IR observables.

The main difference between the approach used in this paper
and previous `hybrid' models (Silva et al. 2005, Lapi et al. 2011) is that in the latter,
the abundance of spheroids as a function of redshift is calculated by means of
a simplified treatment of the merging history of dark matter halos, described by the \cite{1974ApJ...187..425P} formalism.
In this paper, the statistics of spheroidal galaxies are determined purely on a phenomenological
basis, i.e. from their present-day $K$-band LF, with  their backwards evolution describing
their abundance at redshift $z>0$.

\subsection{IR populations}
\label{populations_sec}

Our IR sources division into late-types and proto-spheroids is based
on the results of GPR13. In that paper, the IR luminosity functions (LFs) of
galaxies and active galactic nuclei (AGN) up to $z{\sim}$4 were presented. The
IR LFs were derived by exploiting the deep and extended data-sets of the {\it
Herschel} GTO PACS Evolutionary Probe (PEP, \citealt{2011A&A...532A..90L}).
Different cosmological fields at different depths were included, from the
shallow COSMOS field (3$\sigma$~depth of 10.2 mJy at 160~$\mu$m) up to the
pencil-beam GOODS-S field (3$\sigma$~depth of 2.4 mJy at 160~$\mu$m). Far-IR
{\it Herschel} data from the PEP survey (at 70, 100 and 160 $\mu$m) were
exploited together with data in the sub-mm band from the HerMES survey
(\citealt{2010A&A...518L..21O}, at 250, 350, and 500 $\mu$m), and taking
advantage of the extensive multi-band coverage.

In GPR13, the sources were classified on the basis of an accurate SED-fitting by
comparing data with semi-empirical models from different libraries
(\citealt{2007ApJ...663...81P}, \citealt{2009ApJ...692..556R},
\citealt{2010A&A...518L..27G}). According to the best-fit templates,
sources were divided into five different populations: spiral and
starburst galaxies (i.e. galaxies with no evidence of an
active galactic nucleus, and characterized by $<T_{dust}>$${\sim}$20 and
40-50 K, respectively), AGN1, AGN2 and star-forming (SF)-AGN.  The latter class represents the largest fraction
($\sim$48 \%, see also \citealt{2012ApJ...757...13S}) of
the PEP selected sources, and includes systems whose SEDs were fitted
by means of empirical templates of nearby ULIRGs/Seyferts contaninig an obscured or low-luminosity AGN.
However, the obscured/low-luminosity AGN does not dominate the
energetic budget in these sources; its presence is
detected mostly in the
 narrow mid-IR range, where the flux of the host galaxy presents a minimum.
Spiral
 galaxies constitute the second more numerous class ($\sim$38 \%), followed
by  starburst galaxies representing $\sim$7 \% of the far-IR selected
sources, and lastly by the AGN1 and AGN2 (only 3 and 4 \%, respectively).

Given the large fraction of sources presenting a SF-AGN SED and the large heterogeneity of the templates in this category, SF-AGNs in
GPR13 were further divided in two classes, depending
on their far-IR/near-IR colours (e.g. $S_{100}/S_{1.6}$):
specifically, galaxies fitted by cold Syefert2/1.8 templates were
classified as SF-AGN (Spiral) with far-IR/near-IR colour typical of Spiral/late
type sources, while sources showing the presence of a warmer dust component,
such as NGC 6240, were classified as SF-AGN (SB), with colours typical of
starburst galaxies.

In Fig. \ref{number_density_fig} the comoving number density of the
different populations has been reported, taken from Fig. 18 of
GPR13 (top panel). The number density in GPR13 were computed
by integrating the modified Schechter functions that best reproduced the
different populations down to luminosity $L\sim 10^8 L_{\odot}$. As
already noted by GPR13, the different populations behaves in significantly different and rather opposite ways: while the comoving number density of
spiral and SF-AGN(spiral) strongly increases
 going from the high-$z$ to the local Universe, the number density of starbursts, SF-AGN(SB), AGN1
and AGN2 sources significantly decreases with decreasing redshift.

In detail, spiral galaxies dominate the global density of galaxies in the local
Universe (followed by the SF-AGN(Spiral) systems), but their number density at $z{\sim}1$ already drops by
nearly one order of magnitude.
Starbursts, SF-AGN(SB), AGN1 and AGN2, instead, dominate the number
density at $z>1.5$, whereas their
contribution to the local number density is negligible.
 We  caution the reader that the described trends of the number
  densities, in particolar that referred to the spiral galaxies,
  are not in contrast with the observed decrease of the star-formation density
  from $z{\sim}$1 to $z{\sim}$0 (i.e. \citealt{2006ApJ...651..142H}), directly proportional to the IR
  luminosity density (reported in Fig. 18 of
GPR13, lower panel) and not to the number density.

In this work, we interpret the evolution of the IR comoving number density of starbursts, SF-AGN(SB),
AGN1 and AGN2 sources, progressively decreasing from $z\sim 3$ to
$z=0$, as the transition of all these systems to a population of passive objects. In other words, in our scheme,
we regard all these classes of objects as the progenitors of local
spheroids.
In our formalism, the evolution of the SED of this class of objects is taken
into account by means of the physical model described in
Sect.~\ref{proto_sec}.

On the other hand, on the basis of their SED-classification and on the
positive evolution of their comoving number density as a function of the cosmic time, we classify Spiral and SFR-AGN(Spiral) sources
as late-type galaxies. As already introduced in Sect. \ref{model_sec},
for these galaxy classes, we do not consider any physical model, but we adopt the
same parametric LF evolution as reported in  GPR13.
More specifically, to describe the LF evolution of these
populations at different cosmological times,
we consider both the LLF (computed as a modified Schechter function) and
the evolution of the $L_{star}$ and
$\Phi_{star}$ parameters as presented in GPR13.

\subsection{The proto-spheroid model}
\label{proto_sec}

A long-lasting debate is still open on the formation of spheroids.
The dominant merger-driven scenario, besides remarkably successes,
encounters several difficulties in reproducing some basic properties related to the
scaling properties of spheroids and their evolution
(\citealt{1994A&A...288...57M}; \citealt{2006ARA&A..44..141R};
\citealt{2009NCimR..32....1C}; \citealt{2010A&A...523A..13P}).
For this reason, the old idea (i.e. \citealt{1976MNRAS.176...31L})
that large elliptical galaxies formed their stars at high redshift in
a huge burst of star-formation is nowadays being revisited (see \citealt{2013MNRAS.430.2622D}
and references therein).

The model we adopt for proto-spheroids is based on this idea, and
is similar to the classic `monolithic model' proposed
by various authors (e.g., \citealt{1976MNRAS.176...31L}; \citealt{1994A&A...288...57M}).
More detailed descriptions of the model can be found in
CPM08, \cite{2009MNRAS.394.2001S} and \cite{2011A&A...525A..61P}.

The model described here is designed to reproduce an
elliptical galaxy of present-day stellar mass $M_{*}\sim$ 10$^{11}$~M$_\odot$, which corresponds to
the stellar mass at the break of the local early-type stellar mass function
(\citealt{2004ApJ...600..681B}). Its basic assumptions are the following.

In our scheme, proto-spheroid
galaxies form as the result of a rapid collapse of a
homogeneous sphere of primordial composition, generating an
intense star-formation event (see also \citealt{2002NewA....7..227P},
\citealt{2005A&A...434..553P}). After the initial collapse, the galaxy
 is
allowed to evolve as an `open box' into the potential well of a dark matter halo.

The system is assumed to accrete gas via an infall episode; the infall rate as function of time
 $t$ can be expressed as:
\begin{equation}
(\dot{G})_{inf} \propto \exp{-t/\tau_{inf}}.
\end{equation}
The quantity $\tau_{inf}$ determines the timescale of the collapse;
in this paper, we adopt $\tau_{inf}= 0.4$ Gyrs (\citealt{2011A&A...525A..61P}).

The SFR (expressed in $M_{\odot}$/yr) is calculated as:
\begin{equation}
\psi(t) = \nu M_{gas}(t),
\end{equation}
i.e. it is assumed to be proportional to the gas mass $M_{gas}$ via the star formation efficiency $\nu$,
according to the \cite{1959ApJ...129..243S} law.

The star formation is assumed to halt as the energy of the ISM,
heated by stellar winds and supernova (SN) explosions, exceeds the binding energy of the gas.
At this stage, a galactic wind triggers, sweeping away all the residual gas.
From this point on, the galaxy is assumed to evolve passively.

As described in detail in \cite{2009MNRAS.394.2001S},
in our model we assume
a star formation efficiency $\nu= 15$ Gyrs$^{-1}$ and a Salpeter
initial mass function in the mass interval 0.1-100 $M_\odot$.
Under these conditions, the time of the onset of the galactic wind is $\sim 0.7$ Gyrs.

In the model, the evolution of the chemical composition of the ISM is followed
in detail and as a function of time. The model takes into account also how the
interstellar dust composition varies as a function of time, instead of assuming
a non-evolving dust composition similar to that of the Milky Way (see
\citealt{2007ApJ...657..810D}).

 Also, our model includes a detailed treatment of the rate of Type Ia SNe, which
are assumed to originate from white dwarfs in binary systems, as well as a
computation of their energetic feedback to the ISM
(\citealt{2005A&A...434..553P}). These SNe continue exploding even after star
formation has been quenched by the galactic wind. For this reason, they may
have an important role in keeping the galaxy free of gas and passively
evolving. Indeed, these events occur in a medium already hot and rarefied, where
radiative cooling is inefficient and radiative losses are negligible, so that
most of their blast wave energy can be transferred as thermal energy into the
interstellar medium. Actually, in the context of our model, no additional
source of feedback is required to avoid subsequent episodes of star formation.
It is however worth noticing that semi-analytical models, which, at variance with respect to our simplified treatment, do include
a computation of the gas accretion from the intergalactic medium, and in
recent years begun to take into account the effects of SNe Ia
(e.g. \citealt{2013MNRAS.435.3500Y}; \citealt{2014arXiv1402.3296G}), still indicate that `radio-mode' AGN feedback is required in order to avoid low-redshift star formation episodes in ellipticals.

The chemical evolution model for
 proto-spheroids of CPM08 has then been used by
\cite{2009MNRAS.394.2001S} to calculate how the theoretical SED is
expected to vary as a function of time, once the chemical
evolution of interstellar dust is taken into account in a self-consistent way. To this purpose, the spectro-photometric code GRASIL
(\citealt{1998ApJ...509..103S}) has ben used.
The unique feature of GRASIL is that it allows to track the
 luminosity evolution
of composite stellar populations by taking into account the effects of interstellar dust.

In this paper, the spectral evolution of the proto-spheroid population is modelled by means of the theoretical
SEDs calculated by \cite{2009MNRAS.394.2001S}.

 No attempt is made to model composite 
galaxies, i.e. objects including both bulges and disks. 
These systems are particularly complex since they 
may present composite star formation histories 
and geometries for their dust and stellar components in principle different than those assumed in this paper.

\begin{figure}
\centering
\includegraphics[width=80mm,height=160mm]{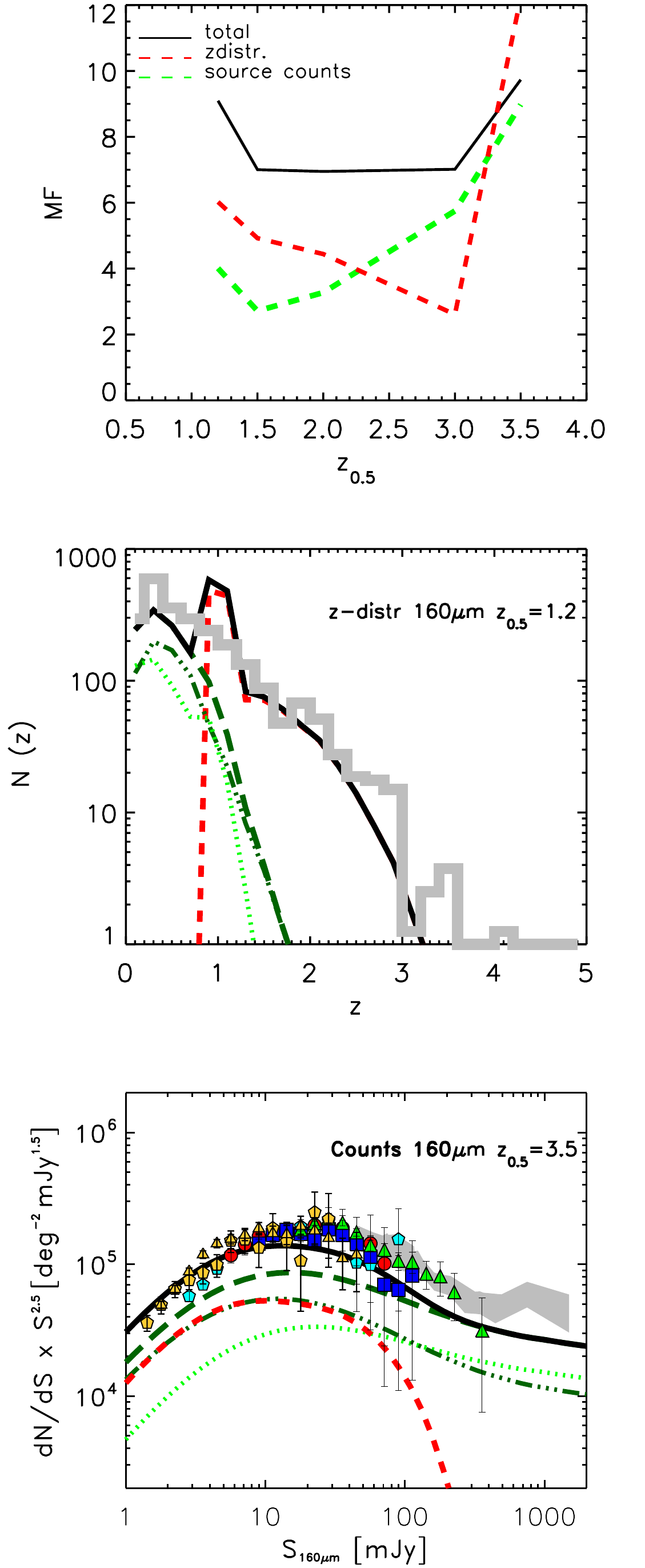}
\caption{
Top panel: 
Merit function as defined in Sect.~\ref{contin_sec} as a function of $z_{0.5}$. 
The red dashed, green dashed and black solid lines are the merit function calculated considering only 
the $K$-band and 160 $\mu$m redshift distributions, the $K$-band and 160 $\mu$m source counts and the four observables altogether, 
respectively. 
Middle panel: 160 $\mu$m redshift distribution obtained considering $z_{0.5}=1.2$. 
Bottom panel: 160$\mu$m source counts obtained considering $z_{0.5}=3.5$.}
\label{chi2}
\end{figure}

\begin{figure}
\centering
\includegraphics[width=80mm]{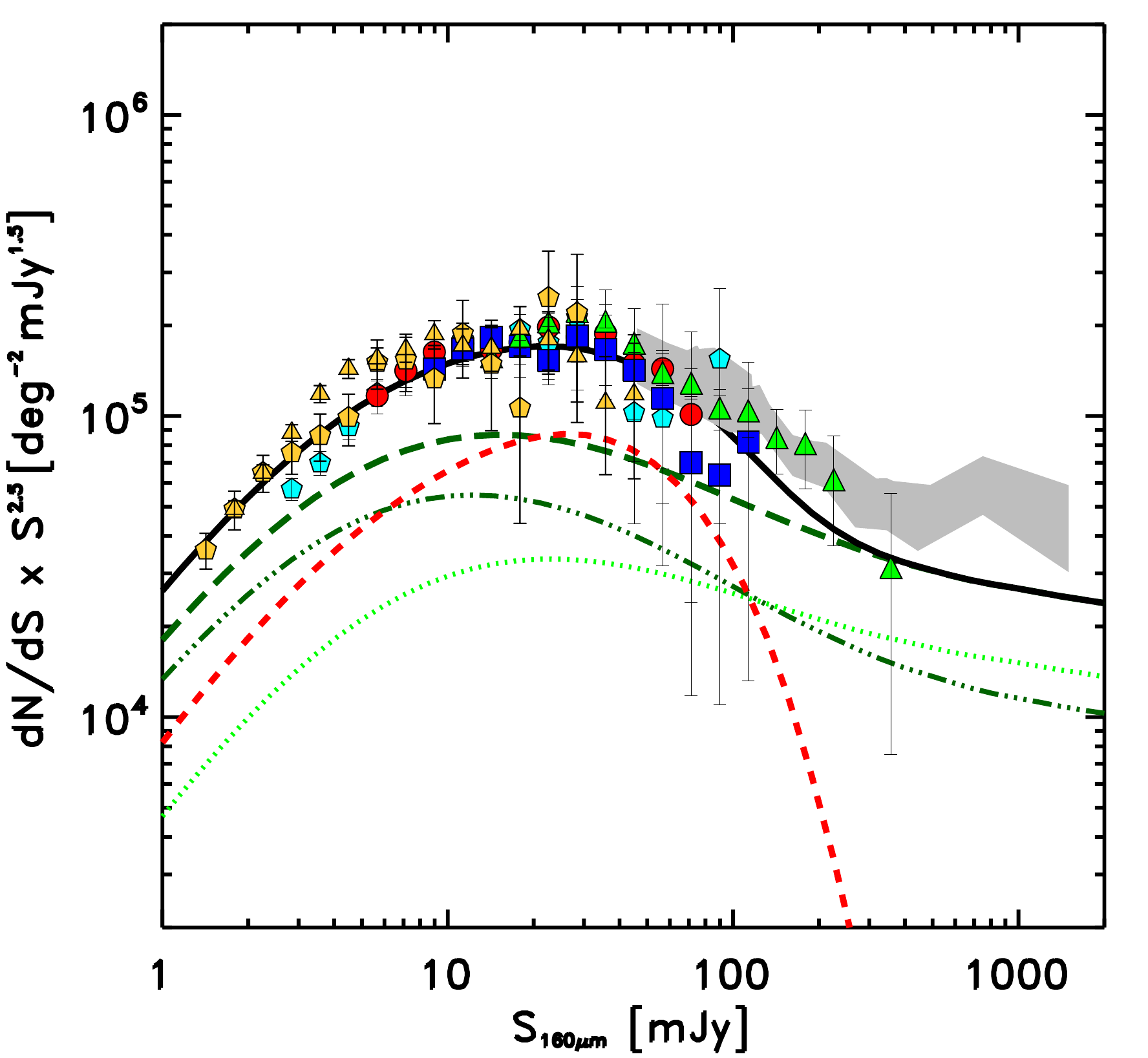}
\includegraphics[width=80mm]{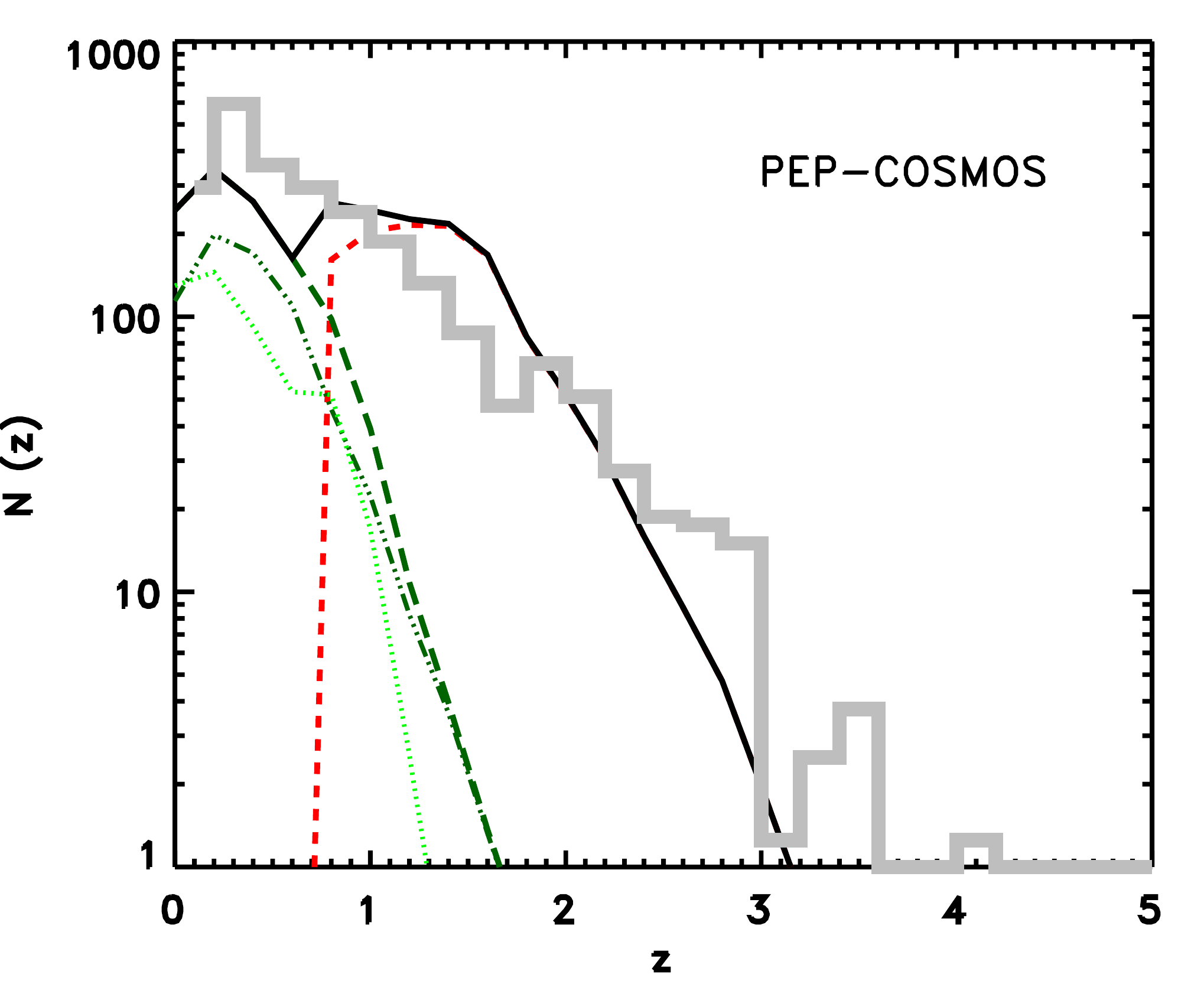}
\caption{Upper panel: 160~${\mu}$m differential extragalactic source counts normalized to the
  Euclidean slope. {\it Herschel} data
  from PEP
 (\citealt{2011A&A...532A..49B}) are represented by red
   filled circles (GOODS-N), blue filled squares (Lockman Hole),
   green triangles (COSMOS) and cyan pentagons (GOODS-S). The grey
   shaded area represents the $1-\sigma$ uncertainty region of the {\it Spitzer} data
   (\citeauthor{2004ApJS..154...93D} 2004; \citeauthor{2006AJ....131..250F} 2006;
  \citeauthor{2010A&A...512A..78B} 2010). The {\it Herschel} data from
  the combined PEP/GOODS-H map (\citealt{2013A&A...553A.132M}) are represented by yellow dots
  (GOODS-S ultradeep) and yellow pentagons (GOODS-N/S deep). Lower panel: Redshift
  distribution of the PEP sources with 160~$\mu$m flux $>$20
  mJy (grey histogram, bin=0.2). The model
 curves for the different populations are plotted in different colours (black solid: total;
 dotted green: spiral; dot-dot-dot-dashed dark-green: SF-AGN(Spiral), long-dashed
 dark-green: spiral+SF-AGN(Spiral); dashed red: proto-spheroids).
 For the model results, a piecewise constant continuous formation for proto-spheroids has been assumed
 as described in Sect.~\ref{contin_sec}. }
\label{counts_fir_fig}
\end{figure}

\begin{figure}
\includegraphics[width=80mm]{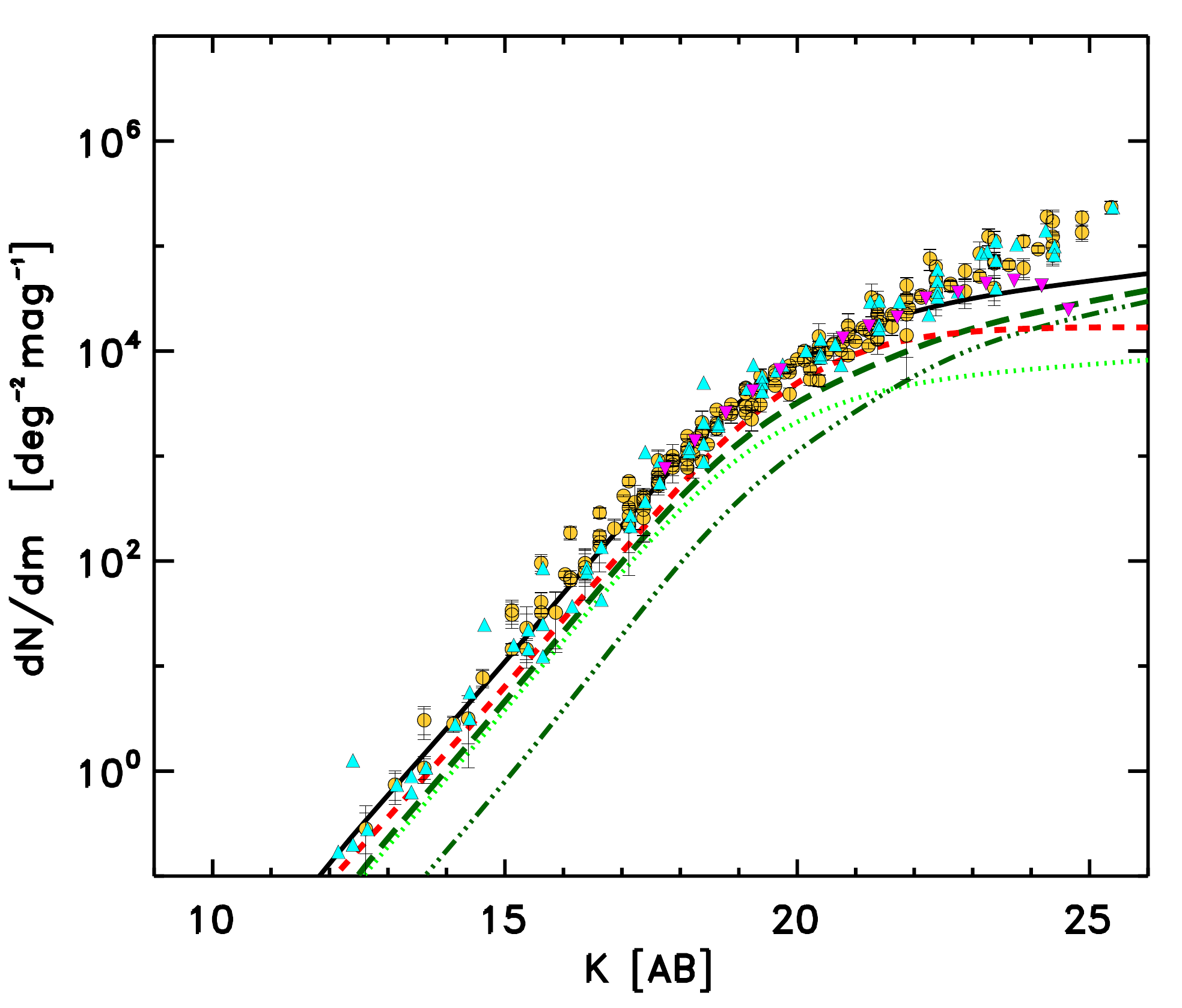}
\includegraphics[width=80mm]{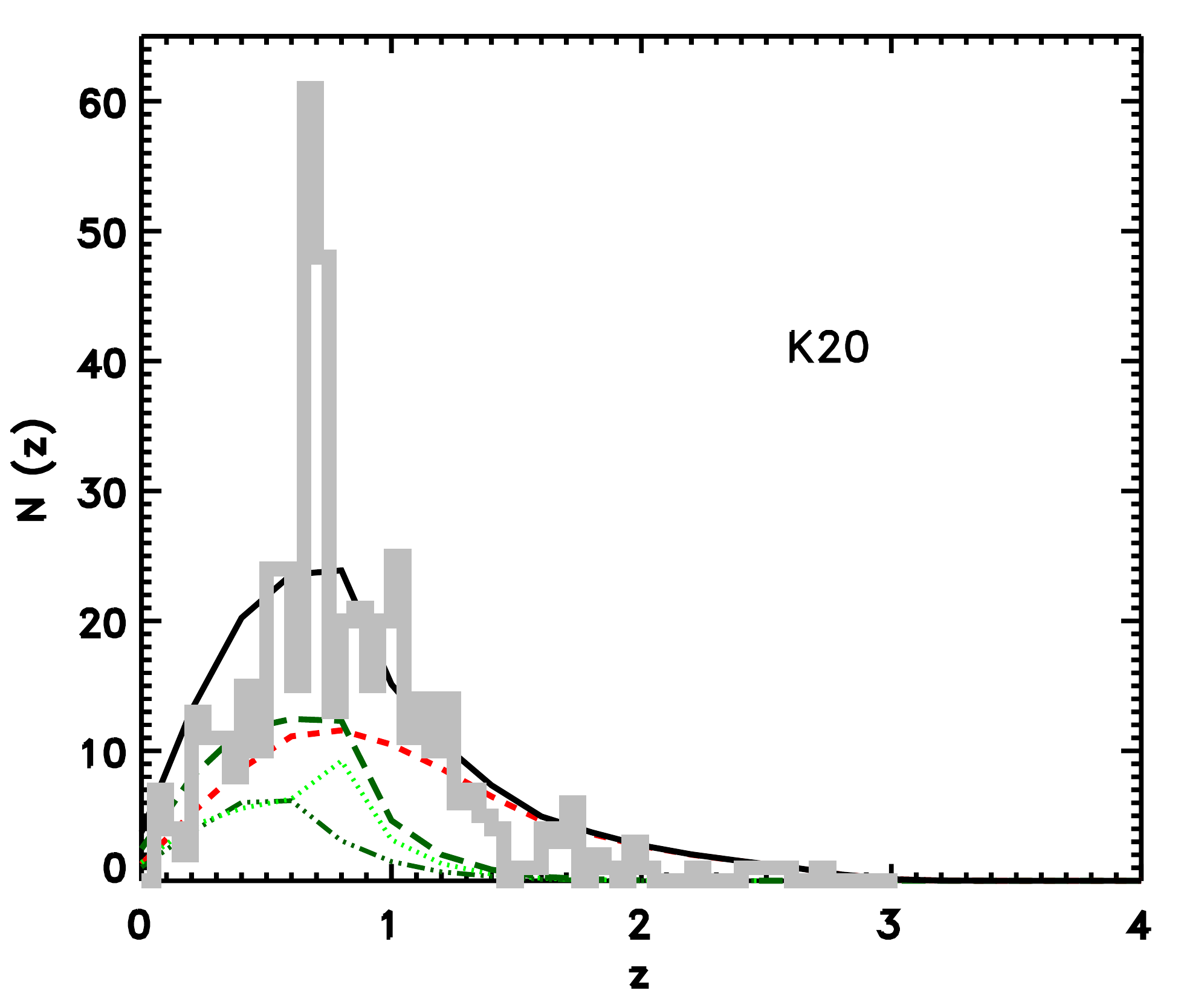}
\caption{Upper panel: Observed and predicted $K$-band differential extragalactic source counts. Observational data are from the Durham public
   compilation (website:   http://star-www.dur.ac.uk/~nm/pubhtml/counts), from
   \citet{2000MNRAS.312L...9M} and from the UltraVISTA survey
   (95\% complete to $K$=23.8, \citealt{2012A&A...544A.156M}), and are shown as yellow circles, cyan triangles and
   pink triangles, respectively.
   The curves represent the same model predictions as in
   Fig. \ref{counts_fir_fig}. Lower panel: Redshift distribution of the $K$-band selected sources from
  two surveys. {\it Top}: K20 survey
  (\citealt{2005A&A...437..883M}) at a magnitude limit of
  $K{\sim}21.9$ ($K[Vega]=20$)  and over an area of 55~arcmin$^2$
  (bin=0.05).  The curves represent the model predictions shown in Fig. \ref{counts_fir_fig}. }
\label{counts_kband_fig}
\end{figure}

\begin{figure}
\centering
\includegraphics[width=80mm]{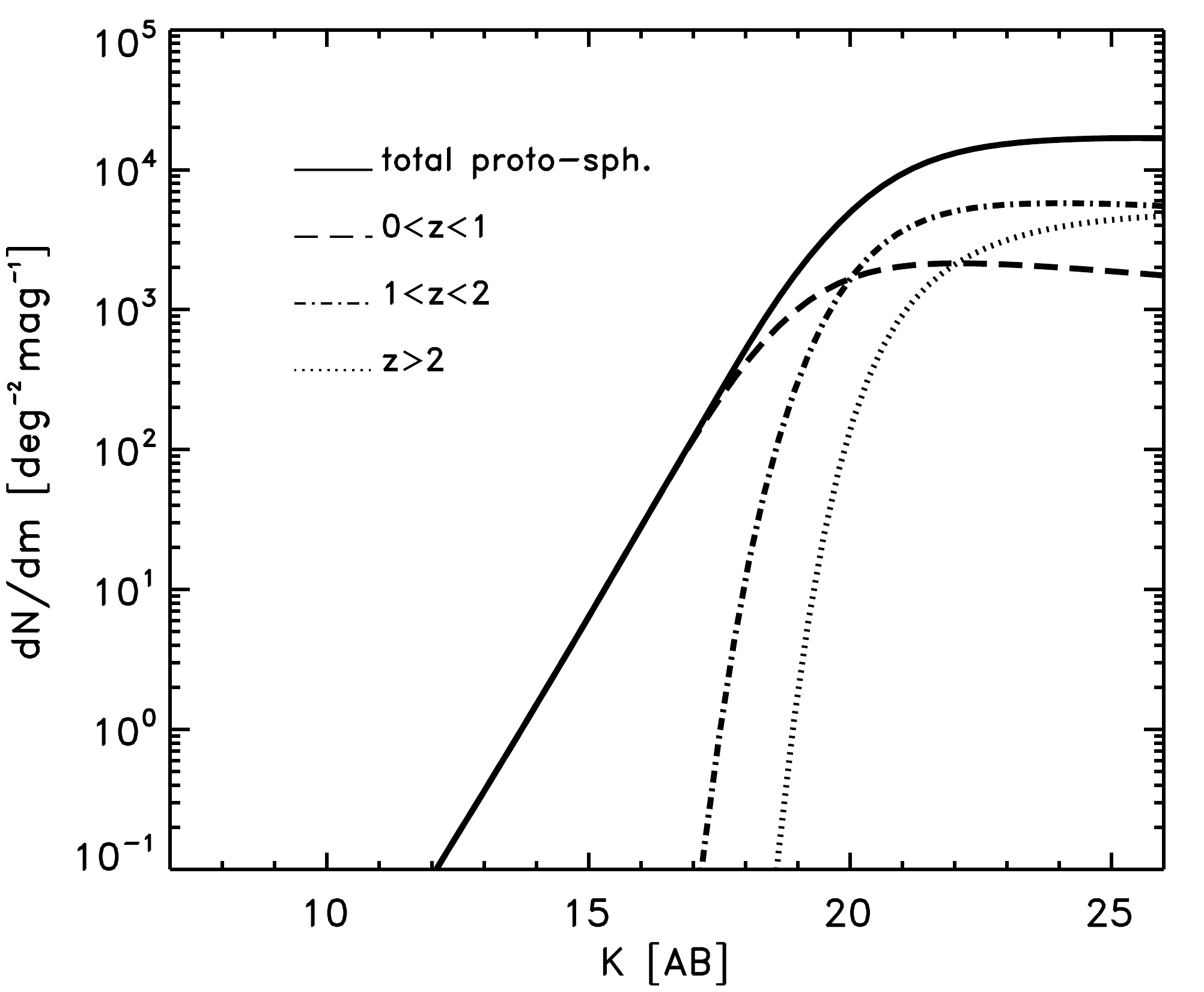}
\includegraphics[width=80mm]{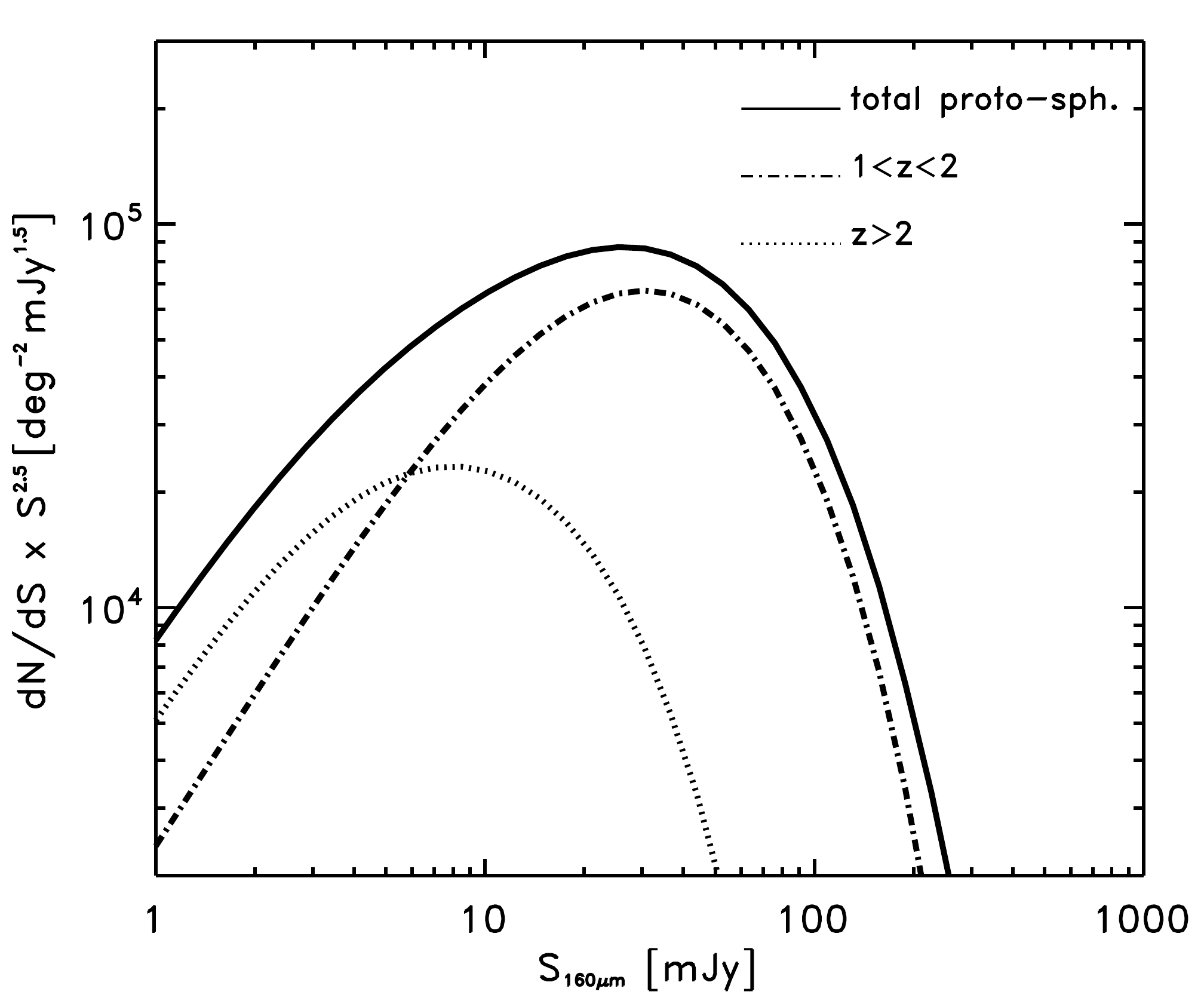}
\caption{Theoretical differential proto-spheroids source counts in
  various redshift bins. Upper panel: $K$-band counts. Lower panel: 160~$\mu$m counts (normalised to the
  Euclidean slope). A piecewise constant continuous formation for proto-spheroids has been assumed
 as described in Sect.~\ref{contin_sec}.}
\label{countsz_fig}
\end{figure}


\section{Results }
\label{results_sec}

In this section, we present the model predictions (in terms of source
counts, redshift-distribution and luminosity function), together with
the corresponding observables.
As reference bands we choose the $K$-band,
probing the old stellar populations, and the far-IR 160~$\mu$m band,
probing the still-forming stellar populations.

For the far-IR we use as reference the PEP survey. In this survey, sources are
selected at 100~$\mu$m and 160~$\mu$m (also at 70 $\mu$m, but only in the GOODS-S field).
We consider the 160~$\mu$m sample as our reference sample, to allow a
direct comparison with the results of GPR13, also based on the
160~$\mu$m PEP catalogue. Moreover, the 160~$\mu$m sample
reaches slightly higher redshifts.

As already mentioned in Sect.~\ref{model_sec}, we normalize the number density of late-type galaxies according to the
IR LF as determined by GPR13.
The LF in the $K$-band as a function of redshift is calculated on the basis of
appropriate colour transformations from the best-fitting SED templates.

At $z=0$, the assumed LF of spheroids is the local luminosity function of early-type galaxies from
\citet{2001ApJ...560..566K};  at redshift greater than zero and in the IR, the proto-spheroid LF is
determined on the basis of the evolution of their SED, calculated self-consistently
by means of the chemo-spectrophotometric model described in Sect.~\ref{proto_sec}.
No evolution is assumed for the faint-end
slope $\alpha$.
At redshift greater than $z_s$, corresponding to the beginning of the proto-spheroids formation epoch, the normalisation is set to zero.

\subsection{Source counts and redshift distributions}
\label{counts_sec}

At any given flux $S^{\star}$, we calculate the differential number
counts (per unit solid angle) of proto-spheroids galaxies from the integral:\\

\begin{equation}
\frac{dN}{dSd\Omega}(S^{\star})=\int_{0}^{z_{max}}\Phi[L(S^{\star},z),z]\frac{dL(S^{\star},z)}{dS}\frac{dV}{dzd\Omega}{dz}
\label{eqn1}
\end{equation}

\noindent where ${\Phi}(L,z)$ is the luminosity
function calculated at the redshift $z$, $dV/dzd\Omega$ is the
comoving volume per unit of solid angle and $z_{max}$ is the
  maximum redshift where the proto-spheroids LF is sampled
  (i.e. corresponding to the formation redshift.)
The relation between the rest-frame luminosity $L$ and the observed flux $S^{\star}$ at a given redshift $z$ is:

\begin{equation}
logL(S^{\star},z)=logS^{\star}+2 \log\ D_{L} + const+ E(z) + K(z),
\end{equation}
where $D_L$ is the luminosity distance, $E(z)$ is the evolutionary
correction, $K(z)$ is the K-correction and $const$ a constant term
(\cite{1997A&AS..122..399P}; \cite{1998ApJ...501..578S}).

Similarly, we calculate the source redshift distribution (i.e. number
of sources above a given flux limit $S_{lim}$, in a redshift range
$z_{1}<z<z_{2}$ and per uniti of solid angle) from the integral:\\

\begin{equation}
N(>S_{lim})=\int_{z_{1}}^{z_{2}}\int_{L_{min(S_{lim},z)}}^{\infty}\Phi[L(S,z),z] {dL(S,z) \frac{dV}{dzd\Omega}{dz}}
\label{eqn1}
\end{equation}

\noindent where $L_{min(S_{lim},z)}$ is the rest-frame minimum luminosity
corresponding to the flux limit $S_{lim}$ at a given redshift $z$.

\begin{figure*}
\centering
\includegraphics[width=160mm]{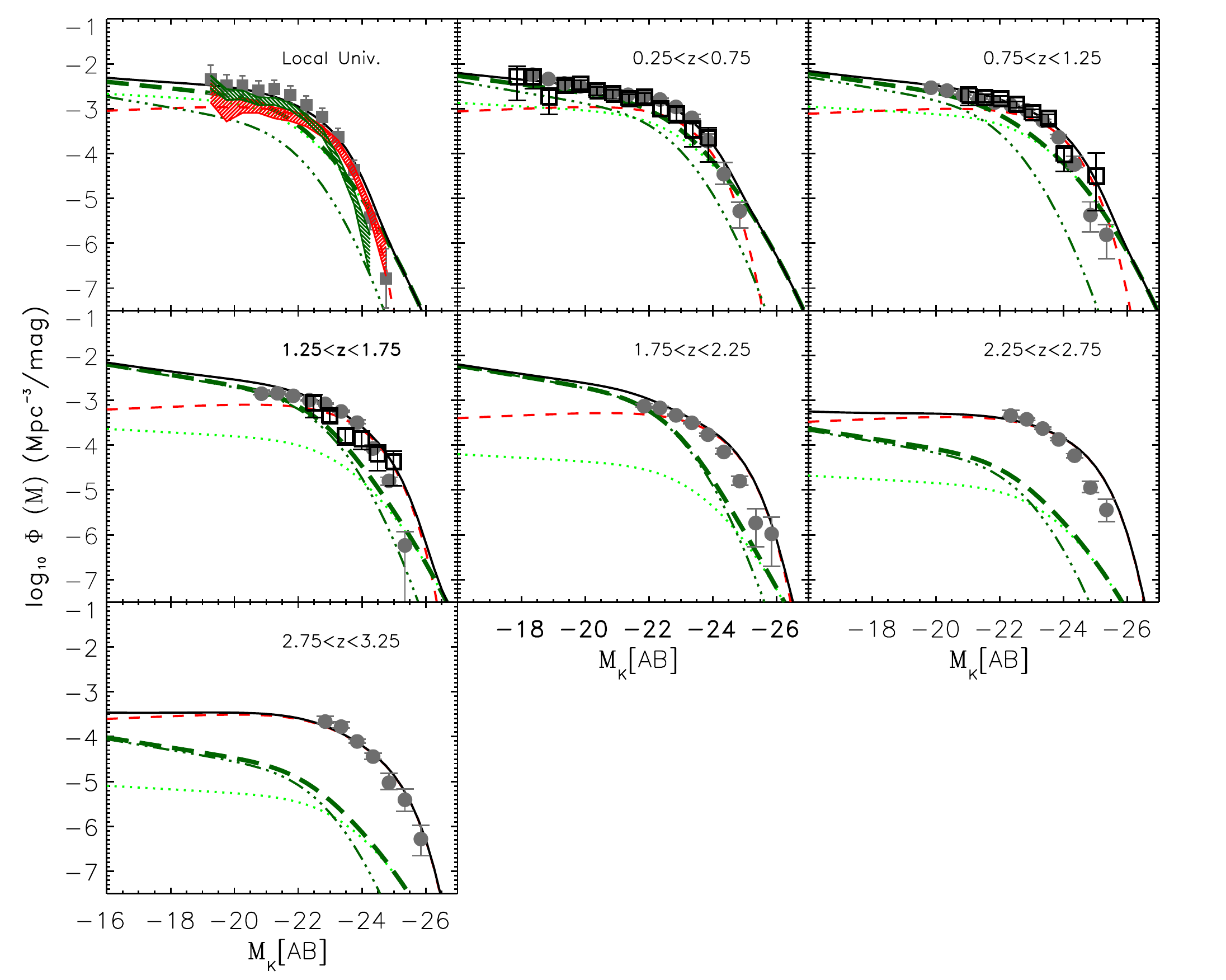}
\caption{
Observed and predicted $K$-band rest-frame luminosity function in eight different redshift
  bins. The model curves for the different populations are plotted in different colours (black solid: total;
 dotted green: spiral; dot-dot-dot-dashed dark-green: SF-AGN(Spiral), long-dashed
 dark-green: spiral+SF-AGN(Spiral); dashed red: proto-spheroids).
 Observational data in the local Universe are from \citealt{2001ApJ...560..566K} (grey squares).
 In the local Universe plot, the red and green shaded areas represent
 the 1-$\sigma$ uncertainty regions of the observed LF for early-type and late-type galaxies, respectively.
 Data points at higher redshift are from
 \citealt{2003A&A...402..837P} (open squares) and from
 \citealt{2010MNRAS.401.1166C} (grey circles).}
\label{lf_fig}
\end{figure*}


\subsection{A continuous formation for proto-spheroids}
\label{contin_sec}

In this work, we assume a continuous formation\footnote{Throughout this work, the proto-spheroids in formation
are those which have started forming stars at a given redshift.} of proto-spheroids, described by 
the function $\frac{\delta N}{\delta z}$ and 
occurring throughout an extended redshift range, which starts at redshift $z_s$ and ending at $z_e$. 

The local number density of spheroids is determined by the local $K$-band LF of \citet{2001ApJ...560..566K}. 
For the sake of simplicity, we assume that $\frac{\delta N}{\delta z}$
is a mass-independent function over a given redshift range $\Delta z=z_s-z_e$. 
By means of this function, 
the global spheroid population is divided in various sub-populations, 
each one born at a different epoch, thus characterised by a different age at the present-day. 
Each sub-population is evolved backwards in luminosity by means of a single spectro-photometric model 
calculated for a galaxy of present-day stellar mass of $\sim 10^{11} M_{\odot}$, corresponding roughly to the break 
of the present-day early type $K$-band LF. 

We believe that $z_e=1$ can be regarded as
a reasonable value for the end of the formation epoch of M${\gsimeq}10^{11}$
M$_{\odot}$ spheroids. This value is consistent with their stellar populations diagnostics,
globally indicating that the bulk of the most-massive early types were already
in place at $z=1$ (\citealt{2006ARA&A..44..141R}). Moreover,
\citet{2013A&A...556A..55I} (see also \citealt{2006A&A...453L..29C}; \citealt{2010A&A...523A..13P}), using the new data from the
UltraVISTA survey (\citealt{2012A&A...544A.156M}) in the COSMOS field,
showed that quiescient  galaxies more massive than $10^{11.2}$ M$_{\odot}$
do not show any evidence of density evolution between $0.8<z< 1.1$
and $0.2 < z < 0.5$.

The assumption of $z_e=1$ is reasonable also in the light of the steep decrease at $z<1$ of
the comoving number density of PEP sources associable to
proto-spheroids, as discussed in Sect.~\ref{populations_sec}.

The assumption that star formation in proto-spheroids is complete at $z \sim 1$ 
prevents our model to account for the presence of Starbursts, AGN1, AGN2 and SF-AGN(SB) objects 
at $z<1$. However, as visible in Fig.~\ref{number_density_fig}, such objects contribute by less than $\sim$ 10\% to 
the number density of IR sources at $0\le z \simlt 1$. 

The quantity $z_s$ is the redshift at which star formation starts, and is set to $z_s=5$. 
We checked that $z_s$ values larger than 5 do not change our results appreciably.

Our aim is to test whether the proto-spheroid formation rate $\frac{\delta N}{\delta z}$ 
can be described by a piecewise constant function.
For the sake of simplicity, we considered the most basic piecewise constant function,
i.e. we divided the redshift range $\Delta z$ in two 
intervals, divided by a value $z_{0.5}$, defined as the redshift at which half of the proto-spheroid population have formed, 
which is the only free parameter of our analysis. 

The quantity $z_{0.5}$ is determined by means of a `merit function', which takes into account the four observables represented by the 
source counts and redshift distribution (at the flux limit of the COSMOS field) at 160 ${\mu}$m and the source counts and redshift distribution 
(at the flux limit of the K20 survey) in the $K$-band. 
This merit function may be intended as a modified reduced $\chi^2$, 
estimated as the sum of the $\chi^2$  obtained for each of the four observables, and where 
the number of degrees of freedom is 
replaced by the number of data bins (flux or redshift bins, depending
on the observable, see \citealt{2009MNRAS.395.2189V}).

The result of our analysis is shown in the top panel of Fig.~\ref{chi2},  
where the merit function is plotted as a function of the quantity $z_{0.5}$ in various separate cases, 
in which the counts and 
the redshift distributions have been regarded as single observables, and in the case where 
these quantities have been considered altogether. 

When the redshift distributions and the source counts are 
considered as single observable, the minimum of the reduced 
$\chi^2$ falls at $z_{0.5}=3.0$ and at $z_{0.5}=1.5$, respectively. 
On the other hand, if the four observables are considered altogether,  
our analysis indicates a minimum of the merit function extending over a region 
spanning from $z_{0.5}=1.5$ and $z_{0.5}=3.0$. 

In the remainder of this work, we choose for the redshift at which half of the proto-spherids have form 
$z_{0.5} \sim 2$ as fiducial value, i.e. a value approximately in-between 
the extremes of the enclosed between $z_{0.5}=1.5$ and $z_{0.5}=3.0$. 

The assumption of a value for $z_{0.5}$ beyond the extremes of the 
distribution causes either a poor description of the 160 $\mu$m
redshift distribution (middle panel of Fig.~\ref{chi2}, considering $z_{0.5}=1.2$), or a significant underestimation 
of the 160 $\mu$m counts at bright fluxes ($S_{160 \, \mu m} \ge 30$
mJy, lower panel of Fig.~\ref{chi2}, considering $z_{0.5}=3.5$). 

To summarize, we assume that half of the proto-spheroids form, at a 
constant rate, over the redshift range 
$1<z <  z_{0.5}$, and half over the redshift interval $z_{0.5} \le z<5$. 

All the results shown in the remainder of this work have been obtained 
with our fiducial assumption, i.e. with $z_{0.5} \sim 2$.

\subsection{Model results versus observations} 
\label{sec_data}
In Fig. ~\ref{counts_fir_fig}, the observed differential far-IR counts calculated at 160 $\mu m$
(upper panel) and the corresponding redshift distribution for the PEP-COSMOS data (lower panel)
are shown, together with the results obtained with our model. 

As shown by the 160~$\mu$m band counts, 
the agreement between data and model
results is remarkably good. In particular, the
proto-spheroid population contributes significantly (being almost half
of the source counts), in the flux range where the differential Euclidean normalized
source counts peak ($S_{160}{\sim}20-30$ mJy).

Also the observed redshift distribution computed at the COSMOS flux limit of
20 mJy (Fig. \ref{counts_fir_fig}, bottom panel) is reproduced by our model.
As discussed by GPR13, at this flux limit, the completeness is close
to 100 \%.


In Fig. \ref{counts_kband_fig},
the observed and predicted $K$-band source counts (upper panel) and
redshift distribution (lower panel) are shown, respectively. 
The model $K$-band redshift distribution shown in Fig.~\ref{counts_kband_fig} is compared with the results of the 
K20 survey \footnote{ The K20 survey (area $\sim$52 arcmin$^{2}$) is composed by sources selected in two independent sky regions,
one centred in the $Chandra$ Deep field South (CDFS, $\sim$32.2 arcmin$^{2}$) and the second centered around the quasi-stellar
object 0055-269 ($\sim$19.8 arcmin$^{2}$). More detailed information can be found in Cimatti et al. (2002).} (\citealt{2002A&A...392..395C}, \citealt{2005A&A...437..883M}).

The $K$-band source counts are well reproduced down to $K{\sim}{23}$
(Fig. \ref{counts_kband_fig}, top panel), whereas at fainter magnitudes
they are underpredicted by our model.
This result is satisfactory considering that, for late-type galaxies,
the $K$-band counts are calculated from the best-fitting SEDs and
from the parametric IR LF evolution obtained in GPR13, whereas the SED
evolution of proto-spheroids is computed solely on the basis of the model described
in Sect.~\ref{proto_sec}.

The agreement between our model and data is
evident also from the computed $K$-band source redshift distribution
(Fig. \ref{counts_kband_fig}, bottom panel). The K20 distribution has been reported since it
represents the most complete spectroscopic survey of a $K$-band
selected sample (i.e. spectroscopic completeness of ${\sim}$96 \%,
\citealt{2005A&A...437..883M}).

At z$\ge$1.5, our analysis indicates that nearly all the sources in the redshift distribution are proto-spheroids.
It may be interesting to compare this result with the morphological
classification of the K20 survey (\citealt{2005MNRAS.357..903C}). 
Of $\sim$ 300 objects belonging to the CDFS region,  $\sim$20-25 sources lie at z$\ge$1.5, and a large fraction of them ($>$80 \%) are classified as irregulars and ellipticals
galaxies, whereas a small fraction ($\sim10-20 \%$) are classified as spiral galaxies. 
The irregular and elliptical systems may be identified with our proto-spheroids, caught during their star-forming phase and already passive, respectively. 
Despite the poor statistics of the K20  sample at z$\ge$1.5, we can regard this result as encouranging. 

In the upper panel of Fig. \ref{countsz_fig}, we show
the predicted differential $K$-band counts computed for proto-spheroids in various redshift bins.
As visible also from the redshift
 distribution plot,
the bulk of the proto-spheroids is between
 $z{\sim}$0 and $z{\sim}2$,
with the counts at $K>19$ dominated by spheroids at $z>1$ and
the counts at brighter magnitudes dominated by $z<1$ sources.
Proto-spheroids at high-$z$ ($z>2$) give a significant contribution only at very faint magnitudes
($K{\sim}23-25$).
As shown in the lower panel of Fig. \ref{countsz_fig},
at 160 ${\mu}$m, instead, most of the proto-spheroid population
lies at $z \gsimeq{1}$, peaking between $1<z<2$.

The predicted $K$-band and far-IR proto-spheroids redshift distribution are
 significantly different (i.e. the $K$-band one extending at
$z{<}1$ at odd with the far-IR), since in our model at $z{\sim}1.6$,
the star formation is completed in half of the proto-spheroids
(corresponding to ${\sim}0.7$ Gyrs after $z=2$, see Sect.~\ref{contin_sec});
i.e. at lower redshifts, they will appear as red, passive objects. The redshift at which the star formation
in all spheroids is complete is $z{\sim}0.8$ (considering the
spheroids formed, at the latest times, at $z{\sim}1$). At lower redshifts, the totality
of these galaxies will evolve as passive systems.

Later, in Sect. 3.4, we will discuss some implications of our results
regarding the stellar mass buildup in early-type galaxies.


\subsection{The evolution of the K-band luminosity function}
\label{K_LF}

Fig. \ref{lf_fig} shows the comparison between the
$K$-band luminosity functions observed at various redshifts and the
one derived by our model and assuming a continuous proto-spheroid formation, as
discussed in Sect.~\ref{contin_sec}.

In the local Universe, we show
the observed  total $K$-band LF as determined by \citealt{2001ApJ...560..566K},
together with the LF derived for early and late-type galaxies, represented by the red and green
dashed areas, respectively.
At higher redshift, the data are taken from the K20 survey up to $z\sim{1.75}$
(\citealt{2003A&A...402..837P}), and from the UKIDSS Ultra Deep Survey
($K{\sim}$23 over 0.7 deg$^{2}$, \citealt{2010MNRAS.401.1166C}) in the
redshift range ${0.25}<z<{3}$.


\begin{figure}
\centering
\includegraphics[width=80mm,height=150mm]{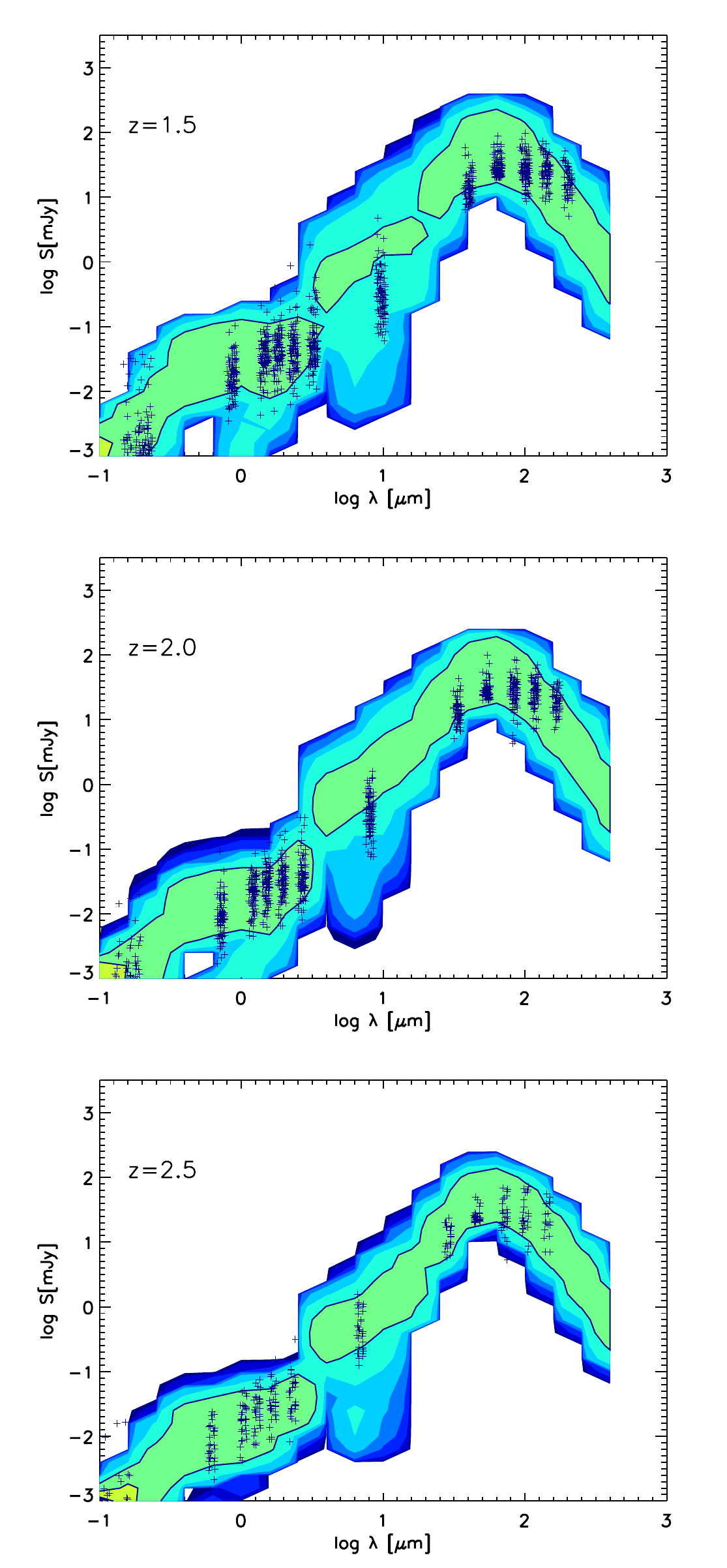}
\caption{
Comparison between the observed and predicted rest-frame SEDs for proto-spheroids in the COSMOS field, calculated at the 160 ${\mu}$m survey flux limit,
and at three different redshifts (from top to bottom: $z=1.5$, $z=2.0$ and $z=2.5$).  
In each panel, the observed SEDs are shown as crosses and represents the sources selected in the redshift range $1.3<z<1.7$, $1.8<z<2.2$ and $2.3<z<2.7$. 
The model SEDs have been calculated as described in Sect.~\ref{sec_seds} and are represented by 
the colour-coded regions, which express the expected number of sources at wavelength $\lambda$ and with flux $S$, normalized 
to the total number of sources at that wavelength . 
The green areas represent regions where the probability of finding a source is greater than 10 $\%$.}
\label{seds}
\end{figure}


\begin{figure}
\centering

\includegraphics[width=75mm,height=70mm]{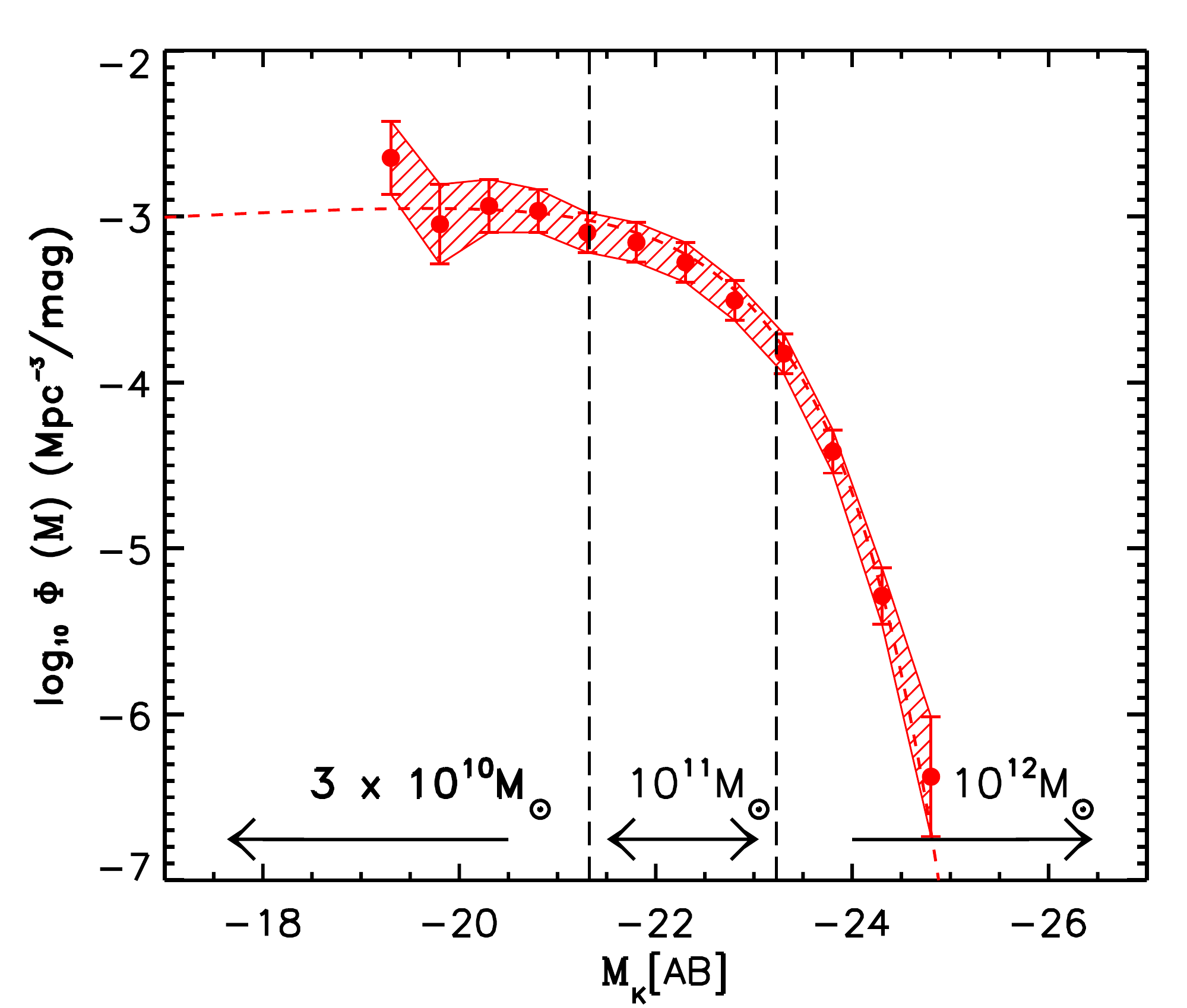}
\caption{The observed local early-type $K$-band LF from
  \cite{2001ApJ...560..566K}. The red short-dashed line and the red shaded area represent the Schechter fit to the 
local $K$-band LF and the associated 1-$\sigma$ uncertainties respectively.
The long-dashed black vertical lines mark the portions of the local LF
associated to the three mass models, as described in Sect.~\ref{downs_sec}}
\label{LLF_ds_fig}
\end{figure}


We find an overall good agreement between the expected model LF
and the data in most of the redshift bins of Fig. \ref{lf_fig}. In the local Universe, the early type LF model estimate is consistent
with the data by construction, since the spheroid LF
has been normalised to the local $K$-band LF. On the other hand, the good agreement between the computed late-type LF and the data
implies that the far-IR selection is not missing a significant number of
sources classified as late-types also in the NIR, and that the SED-fitting procedure used for
the classification is nearly accurate. The slightly shallower
decline, at bright $K$-band magnitudes, of the computed late-type LF with respect to
the observed one reflects the shape of IR-derived luminosity
functions, typically well fitted by a modified Schechter function 
rather than by a pure Schechter (e.g. \citealt{1990MNRAS.242..318S},
\citealt{2004ApJ...609..122P}, \citealt{2012ApJ...758..134S}). 

We find that the late-type population
exhibits a steeper faint-end slope than early-types, and that they dominate
the LF at the fainter luminosities up to $z{\sim}2$.

The largest discrepancy between the model predictions and the data
is visible in the redshift range
$1.75 \lsimeq z \lsimeq 2.75$, where our model overpredicts the number of luminous galaxies, at the bright end of the LF ($M_{K}{\lsimeq}-24.5$).

Possible reasons for the discrepancy are the simplicity of our model
(i.e. we recall that we are using only one proto-spheroid model
for the whole luminosity range, see Sect.~2.2) and the cosmic variance,
whose effects are visible also from the difference between
the bright-end LF of the K20 survey (\citealt{2003A&A...402..837P})
and that of the UKIDSS survey (\citealt{2010MNRAS.401.1166C}).

At high redshift ($2.7 \lesssim z \lesssim  3.25 $), the agreement between model results and the observed LF is remarkable.

Considering the whole redshift range, our results indicate that the contribution of late type galaxies to the total LF is
significant up to $z{\sim}2$. At $z>2$, the $K$-band luminosity function is dominated by proto-spheroids.

\begin{figure*}
\centering
\includegraphics[width=170mm]{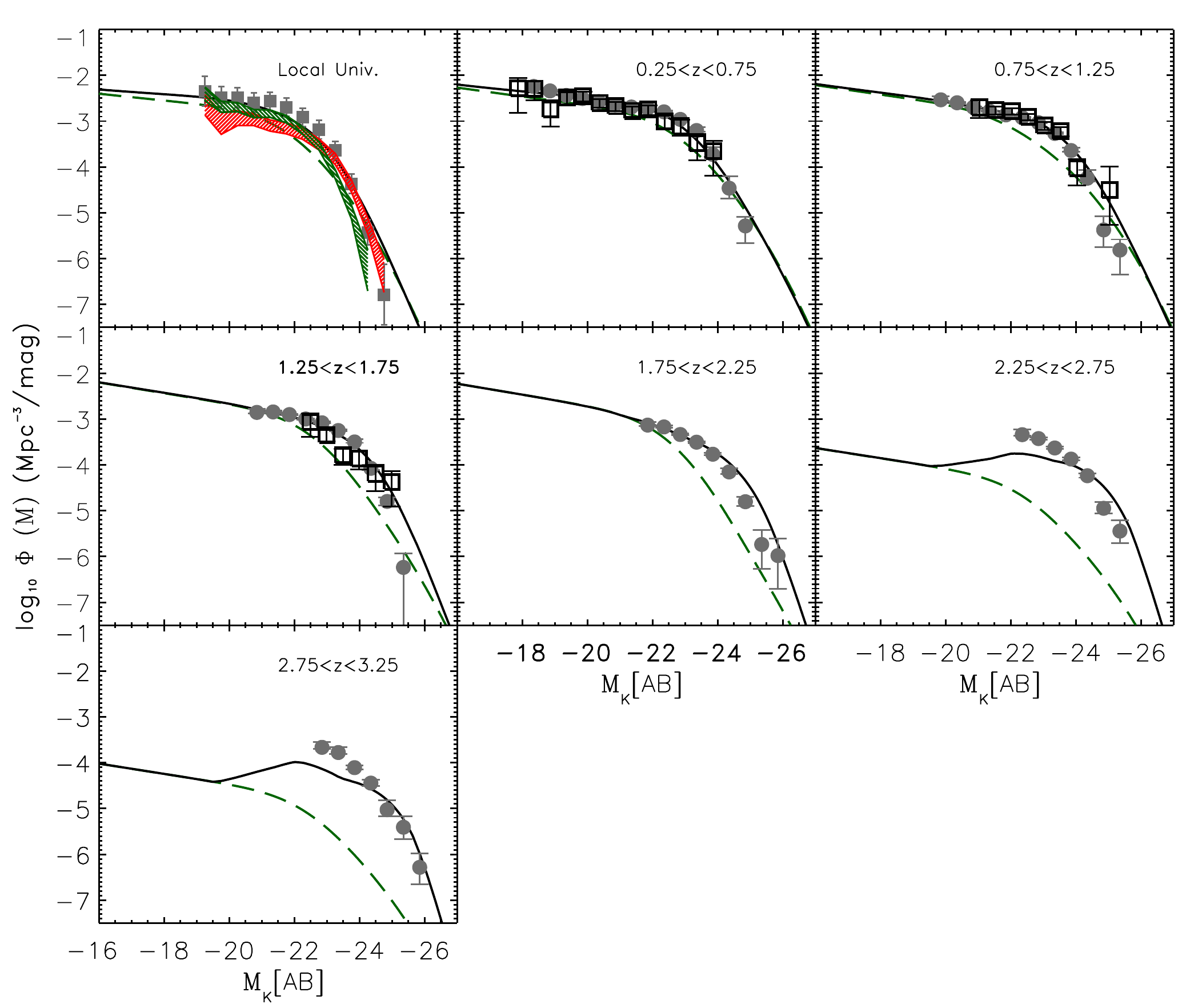}
\caption{Observed and predicted $K$-band rest-frame luminosity function in eight different redshift
  bins. The solid lines represent the total $K$-band LFs 
calculated taking into accounts the effects of downsizing of proto-spheroids, as 
described in Sect.~\ref{downs_sec}. 
The dashed lines represent the contribution of late-type galazies. 
The shaded areas, solid circles and open squares are as described in Fig.~\ref{lf_fig}.}
\label{LF_ds_fig}
\end{figure*}

\begin{figure}
\centering
\includegraphics[width=80mm]{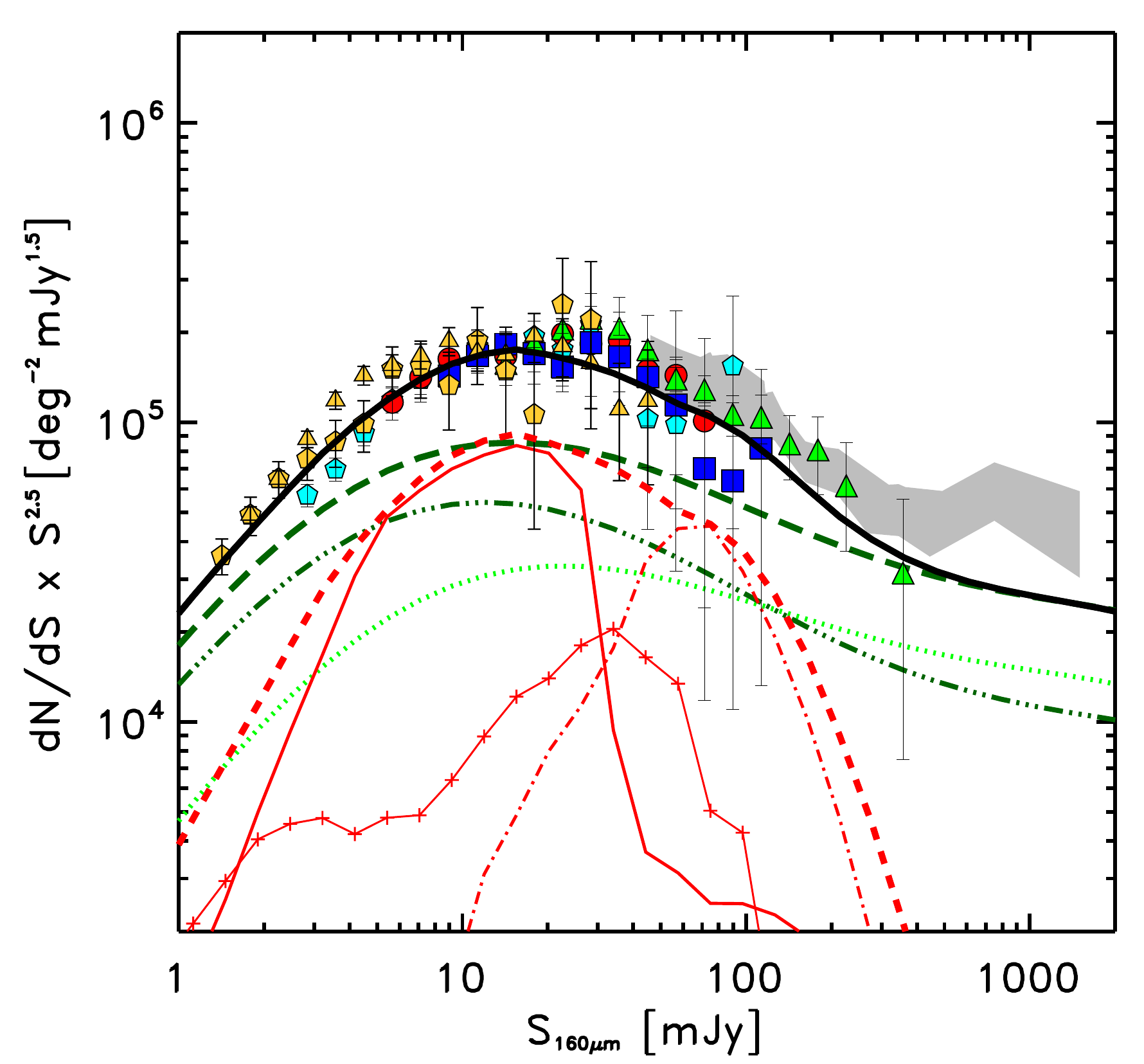}
\includegraphics[width=80mm]{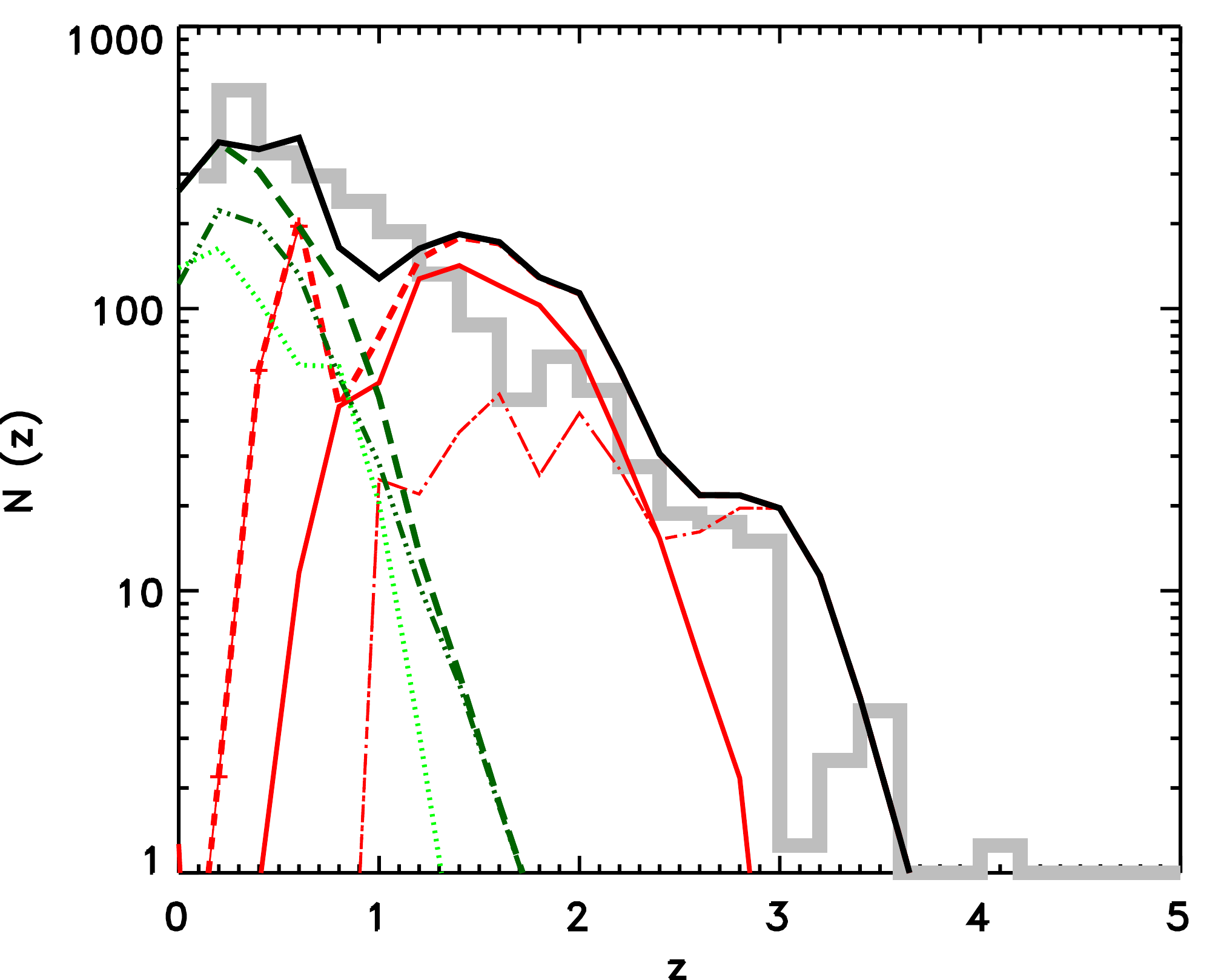}
\caption{
Upper panel: differential extragalactic source counts normalized to the
 Euclidean slope at 160~${\mu}$m. The curves represent the
model predictions, calculated taking into account the downsizing of proto-spheroids.
 The thin solid line with crosses, the thick solid line and the dot-dashed line represent the contribution of the proto-spheroid population of present-day stellar mass $3 \times 10^{10} M_{\odot}$, $10^{10} M_{\odot}$
and $10^{12} M_{\odot}$, respectively; the dotted green line
represents the spiral, the dot-dot-dot-dashed dark-green the
SF-AGN(Spiral) while the  long-dashed dark-green the
spiral+SF-AGN(Spiral) populations. The
observational data are as described in the top panel of Fig.~\ref{counts_fir_fig}.
In the lower panel, the redshift distribution is shown. The curves are as in the top panel and the observational data are as described in the bottom panel of Fig. ~\ref{counts_fir_fig}.}
\label{ds_FIR_counts_fig}
\end{figure}

\begin{figure}
\centering
\includegraphics[width=80mm]{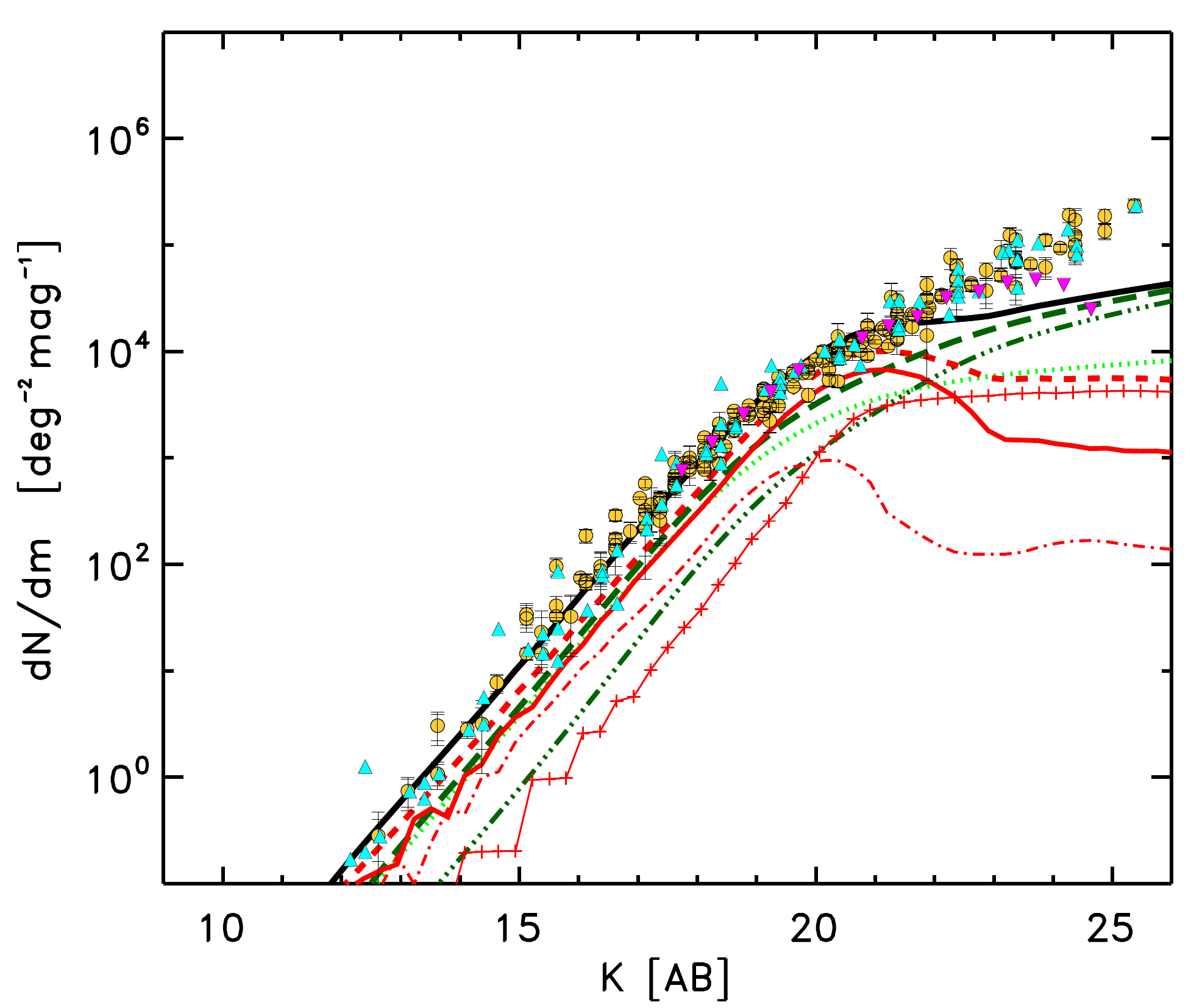}
\includegraphics[width=80mm]{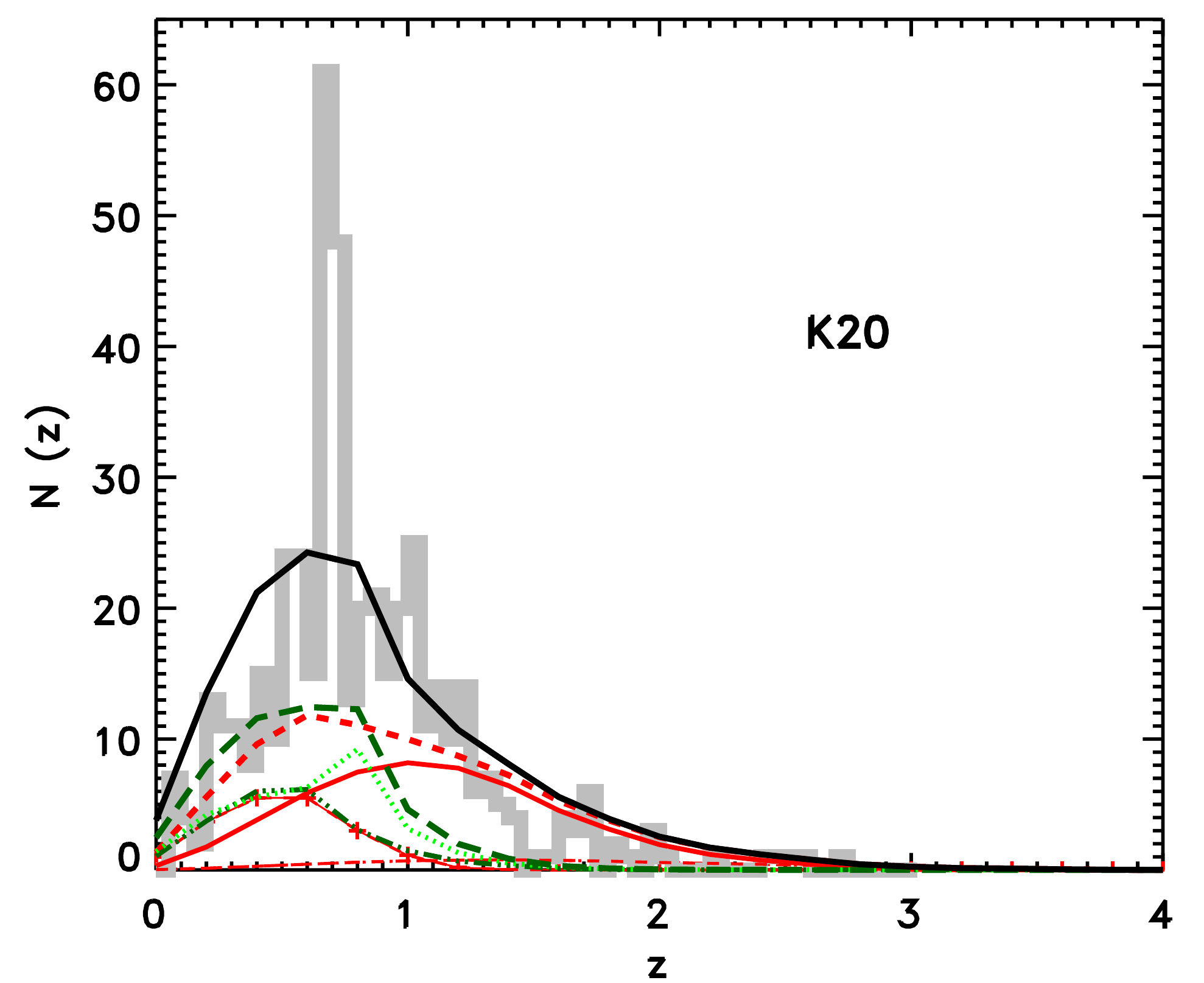}
\caption{
Upper panel: $K$-band differential extragalactic source counts calculated taking into account
the effects of downsizing in proto-spheroids as described in Sect.~\ref{downs_sec}. The curves are as in Fig.~\ref{ds_FIR_counts_fig}. 
The  observational data as described in Fig. ~\ref{counts_kband_fig}.
Lower panel: Predicted redshift distribution of the $K$-band selected sources calculated taking into account the
effects of downsizing. The curves are as above and the observational data are as described in Fig.~\ref{counts_kband_fig}.}
\label{ds_K_counts_fig}
\end{figure}

\subsection{Observed and predicted SEDs}
\label{sec_seds}

As further check, in this Section we aim at testing how the SEDs of our proto-spheroid 
model galaxies compare with the observed SEDs of the non-spiral galaxies of the GPR13 sample. 

In Fig.~\ref{seds}, the theoretical proto-spheroid rest-frame SEDs  are compared with the observed SEDs at high redshift (z$\ge$1.5).
The comparison has been performed in the COSMOS field, considering the 160 $\mu$m flux limit of 20 mJy, at which the completeness of the survey is close to 100 \% (see Sect.~\ref{sec_data}).
The model predictions have been computed at three redshifts  (z=1.5, 2 and 2.5), while the observed SEDs have been selected in the redshift bins 
1.3$<z<$1.7, 1.8$<z<$2.2 and 2.3$<z<$2.7.

The expected model SEDs have been simulated, at each redshift, starting from the theoretical proto-spheroids luminosity function calculated at 160 $\mu$m. 
By construction, the total proto-spheroids LF at a given redshift $z$ is the sum of LF of the proto-spheroid sub-populations formed at different epochs and 
present at that redshift.  
At each luminosity and at each redshift, we can define a weight $w_i$ relative to an extant sub-population, given by the ratio between its LF and the total LF 
at that redshift. 
The SEDs plotted in Fig.~\ref{seds} are those of the extant $i-$th sub-populations at $z=1.5$, $z=2$ and $z=2.5$, each multiplied by its weight $w_i$. 

Overall, the theoretical SEDs are in satisfactory agreement with the observations of GPR13, in particular at $z=2.5$. 
At lower redshift, the model SEDs are a factor $\sim 5$ larger than the observed ones in the Mid-IR regime, 
indicating a warm-dust temperature higher that that inferred from the observed galaxies. 
However, this discrepancy does not affect the main results of this study, which is based on Near-IR and Far-IR data, where the observed SEDs are 
in good agreement with the model results. 

\subsection{The effects of downsizing} 
\label{downs_sec}

The single-mass model described above 
is designed to describe the evolution of the bulk of the population of spheroids.
A more sophisticated treatment should include proto-spheroid models of various masses, 
suited to study the differential evolution of populations of various masses, including their downsizing behaviour. 
To this purpose, we use chemo-spectrophotometric models of proto-spheroids of three different present-day stellar masses, 
and we investigate the effects of downsizing on the main observables considered in this paper. 

The three models are characterised by present-day stellar masses of $3 \times 10^{10} M_{\odot}$, $10^{11} M_{\odot}$ 
and $10^{12} M_{\odot}$. The main parameters describing the models are
the star formation efficiency (from 5 up to 20 Gyrs$^{-1}$
for masses between $3 \times 10^{10}$ and 10$^{12}$~M$_\odot$) and the time of the onset of the
galactic wind (from 0.3 Gyrs up to 0.9 Gyrs). For further details, the reader is referred to 
\cite{2014MNRAS.438.2765C}, where the models are fully described.

 In analogy with the method described in Sect.~\ref{contin_sec}, 
the present-day $K-$band luminosity function of  \citet{2001ApJ...560..566K} is used to normalize 
the total number density of local spheroids. In this case, the present-day early-type $K$-band LF is divided into three parts,  
each one corresponding to a different population. Each population is then evolved backwards by means of the corresponding SED, 
as earlier computed with GRASIL, starting from the star formation history and from the chemical evolution. 
The three models of different masses are associated to the three proto-spheroid populations as follows. 
 The evolution of the faintest galaxies, i.e. those at $K-$band magnitudes M$_K[AB] \simlt 21$, are described 
by the $3 \times 10^{10} M_{\odot}$ model. 
The intermediate mass model, characterised by a present-day stellar mass  of $10^{11} M_{\odot}$, describes 
the evolution of galaxies with magnitudes  $21 \simgt$ M$_K[AB] \simlt 23$. Finally, the evolution of 
the brightest galaxies ($23 \simgt$ M$_K[AB]$) is described by the highest-mass model, with present-day stellar mass 
of $10^{12} M_{\odot}$. The stellar masses have been converted to $K-$ band magnitudes by means of the 
mass-to-light ratios $\frac{M}{L_{K}}(t)$, calculated for each model as a function of the age.

In this case, for each population, 
the functional form of the formation rate $\frac{\delta N}{\delta z}$ is not assumed 
a priori; this approach would generate a considerably wide parameter space and 
a significant degeneracy in the results. 

In the last few years, various observational investigations of early-type galaxies 
have been focused on the determinations of the ages of their stellar populations (\citealt{2006ARA&A..44..141R}), 
providing clear relations between present-day stellar masses and formation redshift. 
Here we assume the relation between formation epoch and mass as
determined by \cite{2010A&A...524A..67M}, based on the multiband SED-fitting analysis of a zCOSMOS sample 
of early type galaxies up to $z=1$. 
The work of \cite{2010A&A...524A..67M} present measures of the
lookback time to formation for galaxies of stellar mass between $10^{10} M_{\odot}$ and  $10^{11} M_{\odot}$.

In our picture, for each population the formation rate $\frac{\delta N}{\delta t}$ is given by a gaussian 
function centered at the age measured by \cite{2010A&A...524A..67M}, where the standard deviation corresponds to the associated 
age dispersion. It is worth stressing that the ages measured by \cite{2010A&A...524A..67M} have been converted to our cosmology.

For galaxies of $10^{12} M_{\odot}$, we assume a mean present-day age  of 12 Gyrs, in agreement with other independent studies 
(see, e. g., Renzini 2006), and we 
adopt the same $1-\sigma$ uncertainty as the one of the systems of mass $10^{11} M_{\odot}$. 

The three $K$-band magnitudes used to divide the local $K$-band LF
have been calculated from the 
present-day SEDs of our models, computed at the
mean age of each population, i.e., following \cite{2010A&A...524A..67M}, at 
${\sim}$ 7.8 Gyr, 10.5 Gyr and 12 Gyrs for the
$3 \times 10^{10}M_{\odot}$, $10^{11} M_{\odot}$ and $10^{12}
M_{\odot}$ model, respectively.

In Fig. \ref{LLF_ds_fig}, we show our division of the local $K$-band
LF. The vertical dashed lines mark the extremes of the three intervals
the LF has been divided into. Each line corresponds to a  magnitude value falling half-way between the present-day magnitudes of two contiguous models. 
The total population of spheroids has then been divided into three populations of different masses. In analogy with our single-mass model, 
the backwards evolution of each population is determined by the evolution of the SED of the corresponding model. 

The evolution of the $K$-band LF obtained by taking into account the effects of downsizing is shown in Fig.~\ref{LF_ds_fig}. 
At redshift $z>2$, the observed LF is now slightly underestimated at the break magnitudes, i.e.  at $M_K \sim -22$.
In principle, this discrepancy could be reduced by 
fine-tuning the adopted mean age of the $3 \times 10^{10}$ and $10^{11} M_{\odot}$
proto-spheroids, e. g. by slightly increasing this quantity, in agreement
with the results of other studies of local early type galaxies (see
\citealt{2006ARA&A..44..141R}).
However, we choose to use the age-mass relation of \cite{2010A&A...524A..67M} at face value and to not 
regard this disagreement as particularly worrying, considering the uncertainty of the present-day age of such objects, of the order of 
1 Gyr (\citealt{2010A&A...524A..67M}). 

Another remarkable effect of downsizing in the predicted high-$z$ LF is in the slope of the faint end: 
in our picture, the least massive progenitors are very rare at $z>2$ and this causes a strong 
fall-off of the LF at $M_K>-22$. 

In general, at lower redshift, the predicted total $K-$band LF is in good agreement with the observations. 

The effects of the proto-spheroids of various masses are visible also 
in Fig.~\ref{ds_FIR_counts_fig}, where we show the far-IR differential counts and redshift distribution. 
From the top panel Fig.~\ref{ds_FIR_counts_fig}, we can see that the
total proto-spheroidal contribution is similar to the ones computed with
 the single-mass model, shown in Fig.~\ref{counts_fir_fig}. 

The $10^{11} M_{\odot}$ mass model gives a dominant contribution to the far-IR source counts, 
in particular in the flux range between $\sim 2$ mJy and $\sim 30$ mJy, 
and to the redshift distribution from $z\sim 0.8$ to $z\sim 2$ (lower panel of Fig.~\ref{ds_FIR_counts_fig}). 

It is also worth noting that the introduction of downsizing allows us to reproduce slightly 
better the observed redshift distribution 
of far-IR sources at $z>2.5$, thanks to the contribution of the most massive systems (i.e. $10^{12} M_{\odot}$).

The predicted $K-$band differential counts and the redshift distribution of K-band sources are shown 
in the top and bottom panel of Fig.~\ref{ds_K_counts_fig},
respectively. 
In general, the total counts are very similar to the ones shown in
Fig.~\ref{counts_kband_fig} (upper panel, i. e. assuming a single-mass model to describe 
the whole population of proto-spheroids), with the exception of the faint-end, where 
the disagreement between data and model results is slightly more
marked. Once again, this effect, could be reduced by
fine-tuning the mean age of the $3 \times 10^{10}$  and $10^{11} M_{\odot}$ mass
model (i.e. moving the peak of formation at higher redshift). 

The similarity of the results computed with the single-mass model and
obtained with the three-mass models are in agreement with other previous
chemical evolution studies of early-type galaxies.
In fact, \cite{1988A&A...202...21M} showed that
the chemical enrichment of the intra-cluster medium is dominated
by galaxies with $L \sim L_{*}$.

The $10^{11} M_{\odot}$ mass model dominates the $K$-band proto-spheroids source counts from the brighter magnitudes down to $K{\sim}22$, while at fainter magnitudes 
the low mass populations (i.e. $3 \times 10^{10} M_{\odot}$)
becomes the prevailing one. 

Consistently with the source counts, the $K$-band redshift distribution
(at the limit magnitude of $K{\sim}21.9$)
shows that the proto-spheroids of $10^{11} M_{\odot}$ 
give the main contribution at $z>0.5$; at lower redshifts,
the major contribution is given by  the lowest mass population (i.e. $3 \times
10^{10} M_{\odot}$). 
The highest mass model (i.e. $10^{12} M_{\odot}$) population is always negligible (given the small area of
the K20 survey and their low volume density); in the source counts,
this population gives a contribution only at very bright magnitudes
($K{\lsimeq} 19$). 



It is worth stressing that the formation rate provided by the age
measures of \cite{2010A&A...524A..67M} allow us 
to reproduce the observed distributions with good accuracy: no fine-tuning of any formation rate was required in this case. 

Our main conclusion is that at most magnitudes/fluxes, the main
contributors of the total differential far-IR and $K-$band counts of
proto-spheroids are represented by a population of present-day mass which corresponds  
roughly to the break of the local early-type stellar mass function.

\subsection{Formation redshift distribution and mass assembly}
\label{zform_sec}

\begin{figure}
\includegraphics[width=80mm]{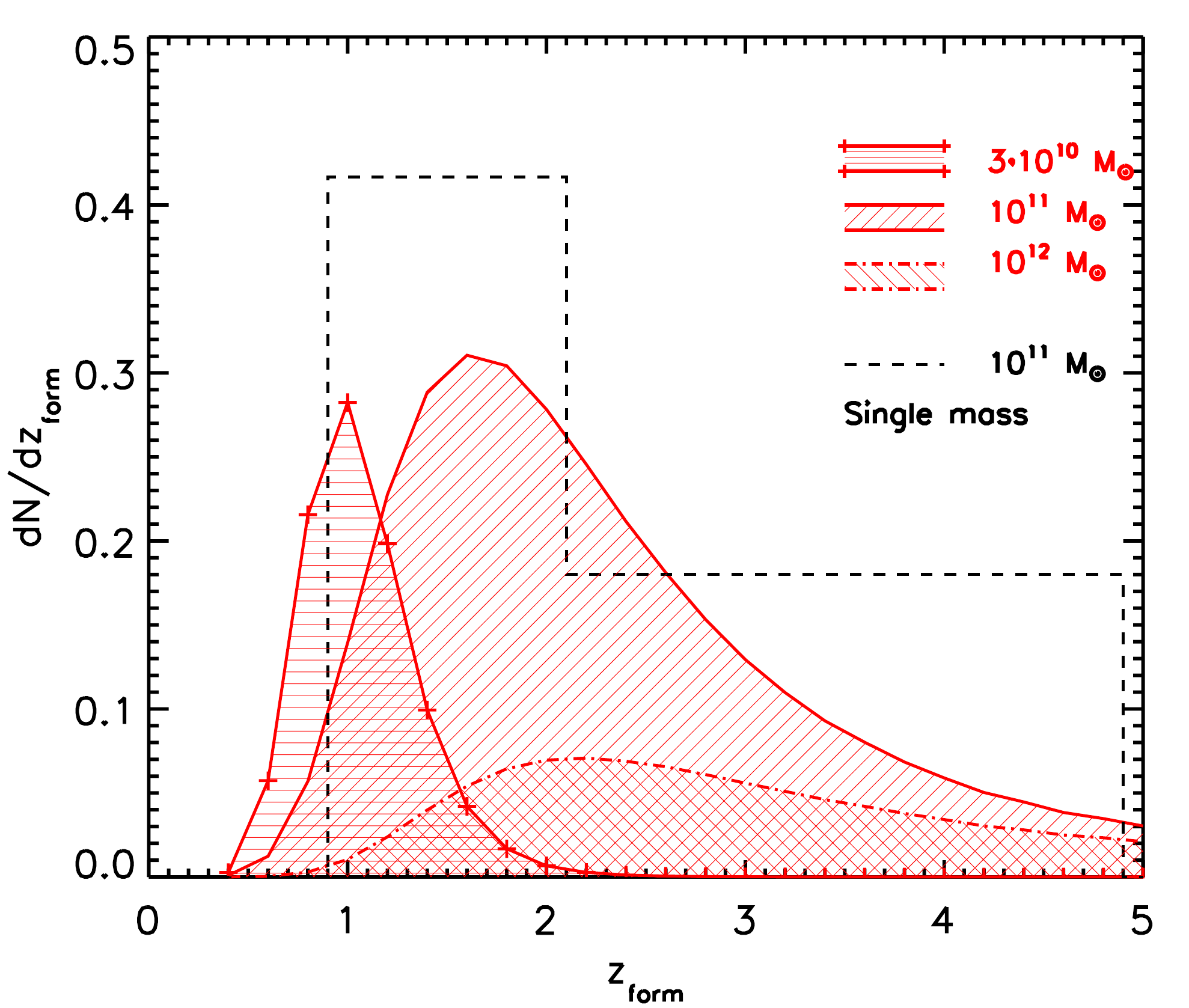}
\caption{Differential proto-spheroid formation rate 
computed in the case of the single-mass model (dashed line), together with 
the formation rates of the three populations (dashed areas). 
Each curve is normalized to the present-day stellar mass density of the corresponding population. 
The curve of the single mass model is normalized to the total
present-day stellar mass in early-type galaxies.}
\label{zformation_fig}
\end{figure}

In Sections ~\ref{contin_sec} and ~\ref{downs_sec} we have shown that it is possible to reproduce a
set of multi-wavelength galactic observables across a wide redshift
range ($0\le z\le 3$) 
by means of a population of late-type galaxies,
whose behaviour is determined empirically by the evolution of their far-IR-LF,
plus a population of proto-spheroids. In Sect. ~\ref{contin_sec}, 
we have assumed a piecewise constant formation rate
across the redshift range $1\le z\le 5$, whereas in Sect.~\ref{downs_sec} we have 
divided the proto-spheroids in three classes of different mass and, for each class, we have assumed 
an observationally-derived age-mass relation. 
In  Fig. \ref{zformation_fig}, we show the evolution of the formation rate 
$dN/dz_{form}$ computed in the case of the single-mass model (dashed line), together with 
the formation rates of the three populations of spheroids described in
Sect.~\ref{downs_sec} (dashed areas).
These latter curves have been normalized according to the local
stellar mass density of the corresponding populations and the sum of their integrals has been
normalized to one as for the single-mass model. 
We find that the combined contributions of low-mass ($3{\times}10^{10}$
M$_{\odot}$) and high-mass ($10^{12}$ M$_{\odot}$) 
spheroid populations sum up to ${\sim}$35\% of the present stellar
mass density (we adopted 10$^{8}$ to 10$^{13}$ M$_{\odot}$ as extremes for the integration).

As visible in the plot,  the $dN/dz_{form}$ function calculated in the
case of  the single-mass model is in fairly good agreement with the 
one corresponding to $M_{*}= 10^{11} M_{\odot}$ of the three-mass
model (i.e. in both cases, at ${z}{\sim}$2 half of the total present-day mass has already formed) and obtained by a completely independent method.  The global
consistency of the two curves, together with the
dominance of the $M_{*}= 10^{11} M_{\odot}$ population at
$z<2.5$ (see Sect. \ref{downs_sec}), outlines the validity of 
our basic assumption, i.e. that the evolution of the whole proto-spheroid population can be
traced by the evolution of galaxies at the break of the present-day early type $K$-band LF. 

The most direct observables suited to study the mass assembly history of spheroids
is the evolution of their stellar mass density (SMD, i.e. \citealt{2003ApJ...587...25D}).
From the theoretical point of view, studying how the stellar mass of galaxies grows with time
is fundamental to assess the role of various processes in determining one of the most basic
galactic parameters, including the roles of mergers and the relative roles of various
quenching mechanisms in driving the galactic star formation histories.

On the other hand, one should bear in mind  that the stellar mass in not a direct
observable as its observational determination depends on a number of model-dependent assumptions,
including the a-priori choice of an initial mass function, a star formation history and a
metallicity: these factors constitute major sources of uncertainty in the
observational determination of the stellar mass in galaxies (see e.g. \citealt{2007A&A...474..443P}; \citealt{2010MNRAS.407..830M}).

To calculate this quantity, we convert the $K$-band LFs as calculated at various redshifts (Sect.~\ref{K_LF})
and shown in Fig. \ref{lf_fig} into stellar mass functions (MFs) by means of 
the mass-to-light ratios $\frac{M}{L_{K}}(t)$ computed at the time $t$ and, at each epoch, by 
properly taking into account the contributions of galaxies born at different epochs. 
 
Our mass-to-light ratios range from $R \sim{0.2} M_{\odot}/L_{\odot,K}$, characterising the end of the SF epoch,
up to a present-day value of $R \sim 1  M_{\odot}/L_{\odot,K}$. Such values are
comparable to those used by other authors (see
e.g. \citealt{2004ApJ...608..742D}; \citealt{2013MNRAS.436.2892N})
in theoretical studies of the stellar populations of early type galaxies.

The SMD is calculated
from the integral of the stellar mass function $\phi(M_{\star},z)$
 over the mass range 10$^{8}$ to 10$^{13}$ M$_{\odot}$, according to:
\begin{equation}
\rho_{*}(z) = \int_{8}^{13} \phi(M'_{*},z) M_{*}'d\,log M_{*}'.
\end{equation}

\begin{figure}
\centering
\includegraphics[width=80mm]{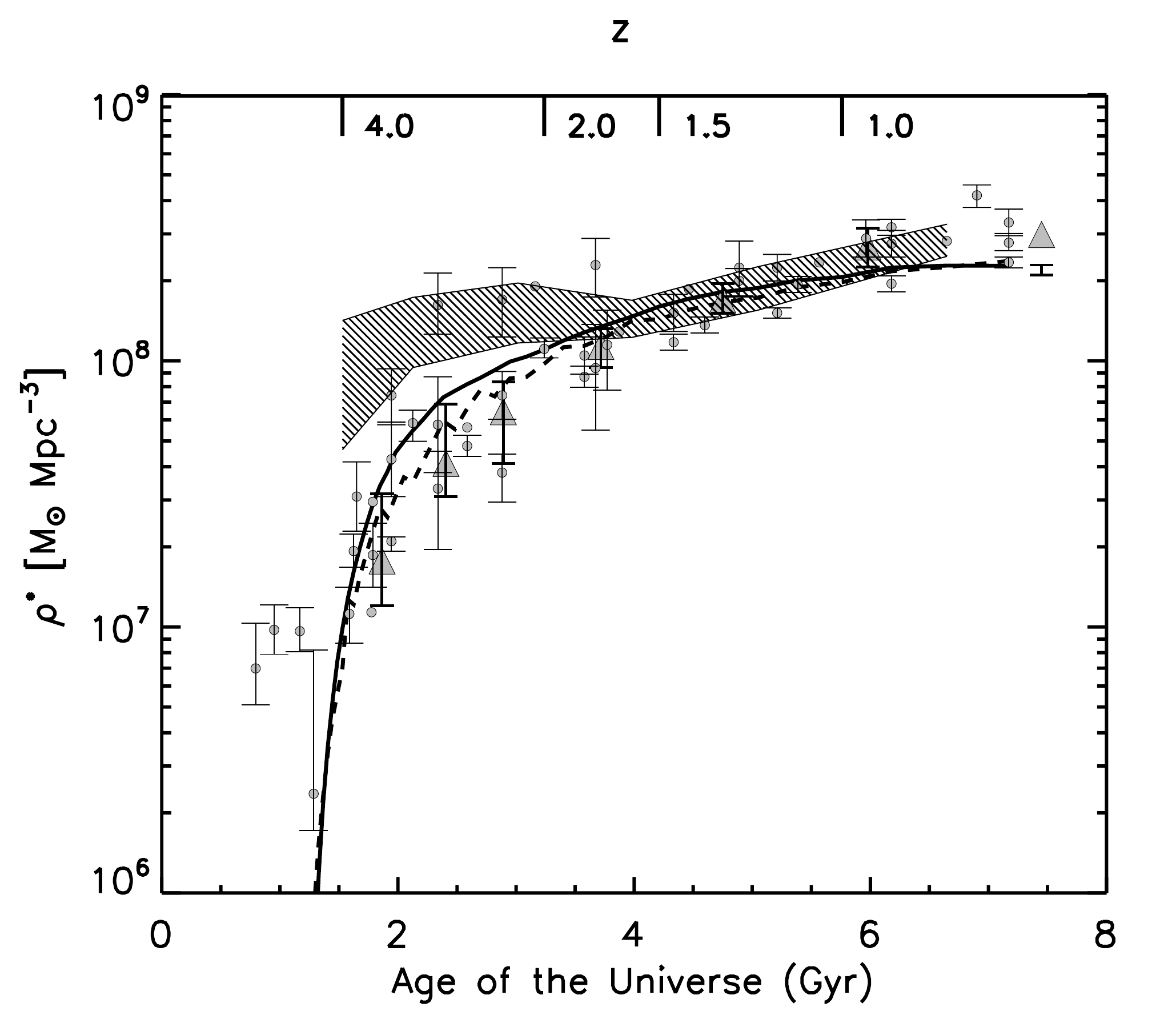}
\caption{Observed and predicted evolution of stellar mass density as a function
of cosmic time.
The solid and dashed lines are the theoretical SMD calculated for
the one-mass and the three-mass models, respectively. The grey circles are the SMD values from the compilation of
\cite{2012A&A...538A..33S}. The  dashed region represents the
estimates from \cite{2012A&A...538A..33S} that take
into account all the glocal sytematic uncertainties. 
The filled grey triangles represent the total SMD as derived in \citet{2013A&A...556A..55I}.}

\label{massdensity_fig}
\end{figure}

In the bottom panel of Fig. \ref{massdensity_fig}, we compare the evolution of the stellar
mass density calculated for spheroids with various estimates from the
literature (see caption of Fig. \ref{massdensity_fig} for details). The SMD from \citet{2013A&A...556A..55I} has been derived in the COSMOS field from
a $K$-band selected sample down to $K$=24 and using the
new UltraVISTA DR1 data release.
In Fig. \ref{massdensity_fig}, we show the total SMD as derived by \cite{2013A&A...556A..55I},
i.e. including both quiescent and late-type systems; the original values have been divided by
$0.55$ to convert them from a \cite{2003ApJ...586L.133C} to a \cite{1955ApJ...121..161S} initial mass function.
In Fig. \ref{massdensity_fig}, we also show a compilation of observational data by \cite{2012A&A...538A..33S}.

In our scenario, at $z>2$ the cosmic star formation is dominated by
proto-spheroids; in this redshift range, the SMD
calculated in this work is consistent with the total estimates
of \cite{2013A&A...556A..55I} and \cite{2012A&A...538A..33S} within the uncertainties.
This result supports a scenario in which most of the mass at high-redshift
(i.e. observed at $z>2$) is in proto-spheroids. Moreover our results
indicate that half of the proto-spheroids mass must have formed at
$z>2$, and the remaining between $z{\sim}1$ and
$z{\sim}2$.

At lower redshift, the SMD values of \citet{2013A&A...556A..55I} and
\cite{2012A&A...538A..33S} are
slightly larger than our estimates; this result is not surprising,
as our estimates are lacking the contribution from late-type galaxies. 

A detailed study of the contribution of late-type systems to the SMD evolution requires
the construction of a library of template SEDs similar to those used in this
paper to study proto-spheroids; such a task is deferred to future work. \\

\section{Conclusion}
In this paper we have interpreted
a set of multi-wavelength observations of galaxies across a large
fraction of the cosmic epoch ($0\le z \le 3$) by means of
a new phenomenological approach.
Our approach matches a `backward' parametric evolution of
late-type galaxies based on the observed IR luminosity functions of
GPR13 to a model for proto-spheroids where
the spectro-photometric evolution of dust has been
calculated self-consistently, on the basis of idealised star formation histories
able to account for a large set
of observational data. These data include the dust budget observed in local passive galaxies and
detected in high-redshift massive starbursts (CPM08; \citealt{2011A&A...525A..61P}) and their SEDs
as observed in both the local and distant universe (\citealt{2009MNRAS.394.2001S}).
This approach has been developed in order to simultaneously
account for observables related mostly to
the star-formation process  and young stellar populations, traced by far-IR emission,
and also for the ones associated to evolved stellar populations, whose presence is detected
in the near-IR. \\
As a first step, following GPR13, IR sources have been divided into late-type and potential progenitors of present-day spheroids
on the basis of their SED and of the evolution of their comoving
number density. The number density of late-type galaxies has been
parameterized as in GPR13. The number density of proto-spheroids has then been normalized according to the local $K$-band LF of early-type galaxies from
\cite{2001ApJ...560..566K}.
Finally, the evolution of their $K$-band and far-IR LFs has been calculated by means of the model SEDs
from the physical chemo-spectrophotometric model described in \cite{2009MNRAS.394.2001S}.
We have also tested the effects of mass downsining on the observables studied in this paper, by dividing the proto-spheroid population 
into three populations of different masses, and tracing their backwards evolution by means of chemo-spectrophotometric models of 
galaxies of present-day stellar mass of $3 \times 10^{10} M_{\odot}$, $10^{11} M_{\odot}$ 
and $10^{12} M_{\odot}$. 
Our main conclusions can be summarised as follows.
\begin{itemize}
\item 
We have considered a continuous formation for proto-spheroids, occurring 
across the redshift range $1\le z \le 5$. In this picture, 
the proto-spheroid formation is described by a  piecewise-constant 
function, where the key parameter is the redshift $z_{0.5}$ at which half the population has formed.  
To constrain this parameter, we have performed a statistical test by
taking into account the following four observables: the 
source counts and redshift distribution 
(at the flux limit of the COSMOS field) at 160 ${\mu}$m and the source counts and redshift distribution 
(at the flux limit of the K20 survey) in the $K$-band. 
Our analysis shows that our full set of observables can be reproduced by assuming $z_{0.5}$ 
in the interval $1.5 \le z_{0.5} \le 3.$
The adoption of values for $z_{0.5}$ lower than 1.5 implies 
a poor description of the 160 $\mu$m redshift distribution; on the other hand, 
the assumption of  $z_{0.5}$ larger than 3 causes a poor description of the 160 $\mu$m counts at the brightest fluxes 
($S_{160 \, \mu m} \ge 30$ mJy). \\

By assuming $z_{0.5} \sim 2$ as fiducial value, we find that 
at the far-IR flux
limit of the PEP-COSMOS survey, all the PEP-selected sources at $z>2$ can be explained as progenitors of local spheroids caught during their formation. 
\item A continuous formation of proto-spheroids with $z_{0.5} \sim 2$  
allows us to reproduce the
differential extragalactic source counts observed at 160~$\mu m$, with
proto-spheroids contributing significantly at the flux range where the differential Euclidean normalized
source counts peak ($S_{160}{\sim}20-30$ mJy).
Also the differential counts in $K$-band are reproduced by our scenario, with the exception of
the faint-end portion. This discrepancy is probably due to the non-evolving
slope of the proto-spheroid luminosity function as adopted in this work. 
\item The differential $K$-band counts computed in various redshift bins indicate that
the bulk of the proto-spheroids lie between $z{\sim}$0 and $z{\sim}2$,
with the counts at $K>19$ dominated by spheroids at $z>1$ and
the counts at brighter magnitudes dominated by $z<1$ sources.
High-$z$ proto-spheroid sources contribute significantly to the $K$-band counts only
at very faint magnitudes ($K{\sim}23-25$).
From the far-IR counts, we have seen that most of the star-forming proto-spheroid population
lies between $1<z<2$. 
As further check, we found a satisfactory agreement between the simulated SEDs of the model proto-spheroids and the 
spectra of non-spiral sources in the COSMOS field at $z>1.5$.
\item Our results obtained by taking into account the effects of downsizing 
show that the main contribution 
to the counts comes from objects of 
present-day stellar mass of $10^{11} M_{\odot}$, corresponding roughly to the break of the local early-type stellar mass function.

\item Finally, we have studied the evolution of the stellar mass density as implied by our
results and we have compared it to other estimates from the
literature. At high-$z$ ($z>2$), we find a good consistency between the SMD
calculated in this work and the total estimates from other authors,
supporting a scenario in which most of the mass at high-redshift
is in proto-spheroids.  Moreover our results
indicate that half of the proto-spheroids mass must have formed at
$z>2$, and the remaining ones between $z{\sim}1$ and
$z{\sim}2$. The same results are obtained also when downsizing is taken into account. 
This confirms the validity of our approach, i.e. that the mass growth
of the proto-spheroid population can be modelled by means of a
single-mass model, describing the evolution of galaxies at the break
of the present-day early-type $K$-band LF.

At lower redshift, the total SMD estimates from literature are
slightly larger than our estimates. This is likely due to the fact that late-type galaxies have not been considered in our study.
 In the future, the contribution of late-type galaxies to
the SMD will  need to be assessed in order to provide more accurate predictions on the evolution of the SMD and
 to gain further hints on their formation history.
\end{itemize}

As a final note, we stress that our results
do not allow us to quantitatively assess the role of
mergers as main drivers of the star formation history of proto-spheroids,
which is probably dominant at large redshifts (\citealt{2009MNRAS.398L..58R}; \citealt{2009MNRAS.394.1956C}). 
 A merger-driven formation of spheroids is still in agreement with our results, but if 
the assembly of the sytems of stellar mass $10^{11} M_{\odot}$  occurs
preferentially at $z>$1.
Presently, the major difficulty of cosmological scenarios, where proto-spheroids
form as the results of a sequence of merging episodes,
is to have the starbursts associated to the most massive systems
completing their star formation histories before $\sim 1$ Gyr, a duration which allows to account for
a wide set of multi-wavelength properties at high redshift as seen here and in other works
(see, e.g., \citealt{2004ApJ...600..580G},
\citealt{2005MNRAS.357.1295S}, \citealt{2011ApJ...742...24L}), as well as their main scaling relations as traced by
the basic stellar population diagnostics of local ellipticals
(\citealt{1994A&A...288...57M}; \citealt{2006ARA&A..44..141R}).

To study the IR properties of galaxies by means of cosmological models, one needs to take
into account the dust production processes self-consistently as in the model used in this work.
To accomplish such a task, the next-generation of $\Lambda-$CDM semi-analytical models
will have to include a detailed
treatment of the chemical evolution of refractory elements (\citealt{2011MNRAS.413L...1C}).

\section*{Acknowledgements}

FP thanks M.Cirasuolo for information on the UKIDSS survey, P.Santini
for sending us her compilation of mass density,  M. Moresco for
sending us his spheroids formation epochs. FP thanks C. Vignali,
P. Andreani, M. Bethermin and I. Davidzon for helpful discussion.
We are grateful to an anonymous referee for valuable suggestions that improved the paper. 
FP and CG acknowledges partial support from the Italian Space Agency (contract
ASI I/005/11/0).
FC and FM acknowledge financial support from PRIN MIUR 2010-2011,
project `The Chemical and Dynamical Evolution of the Milky Way and Local Group Galaxies',
prot. 2010LY5N2T.

\clearpage

\clearpage

\clearpage


\bibliographystyle{apj}
\bibliography{pozzi}

\end{document}